\documentclass[fleqn,usenatbib]{mnras}

\usepackage{newtxtext,newtxmath}

\usepackage[T1]{fontenc}
\usepackage{ae,aecompl}

\usepackage{graphicx}	
\usepackage{amsmath}	
\usepackage{amssymb}	

\newcommand{\beq}{\begin{equation}}
\newcommand{\eeq}{\end{equation}}
\newcommand{\beqn}{\begin{eqnarray}}
\newcommand{\eeqn}{\end{eqnarray}}

\usepackage{color}

\title[MC closure for two-moment radiation transport]
{Monte-Carlo closure for moment-based transport schemes in general relativistic radiation hydrodynamics simulations}

\author[F. Foucart]{
Francois Foucart$^{1,2}$  \thanks{fvfoucart@lbl.gov}\\
$^1$ Lawrence Berkeley National Laboratory, 1 Cyclotron Rd, Berkeley, CA 94720, USA\\
$^2$ Department of Physics, University of New Hampshire, Durham, New Hampshire 03824, USA
}

\date{Accepted XXX. Received YYY; in original form ZZZ}

\pubyear{2017}

\begin{document}
\label{firstpage}
\pagerange{\pageref{firstpage}--\pageref{lastpage}}
\maketitle

\begin{abstract}
General relativistic radiation hydrodynamics simulations are necessary to accurately model a number of astrophysical systems involving black holes and neutron stars. Photon transport plays a crucial role in radiatively dominated accretion disks, while neutrino transport is critical to core-collapse supernovae and to the modeling of electromagnetic transients and nucleosynthesis in neutron star mergers. However, evolving the full Boltzmann equations of radiative transport is extremely expensive. Here, we describe the implementation in the general relativistic SpEC code of a cheaper radiation hydrodynamics method which theoretically converges to a solution of Boltzmann's equation in the limit of infinite numerical resources. The algorithm is based on a gray two-moment scheme, in which we evolve the energy density and momentum density of the radiation. Two-moment schemes require a closure which fills in missing information about the energy spectrum and higher-order moments of the radiation. Instead of the approximate analytical closure currently used in core-collapse and merger simulations, we complement the two-moment scheme with a low-accuracy Monte-Carlo evolution. The Monte-Carlo results can provide any or all of the missing information in the evolution of the moments, as desired by the user. As a first test of our methods, we study a set of idealized problems demonstrating that our algorithm performs significantly better than existing analytical closures. We also discuss the current limitations of our method, in particular open questions regarding the stability of the fully coupled scheme.

\end{abstract}




\section{Introduction}
\label{sec:intro}

Neutrinos and photons play a critical role in numerical studies of astrophysical systems. For example, general relativistic photon transport is required to study radiatively dominated accretion disks, while neutrino-matter interactions are crucial to the explosion mechanism of core-collapse supernovae. In neutron star mergers, neutrinos do not directly affect the dynamics of the merger, yet they are the main source of cooling of the accretion disks and massive neutron stars formed as a result of many mergers. 
Neutrino-matter interactions also determine the evolution of the composition of the matter ejected by these mergers. These neutron-rich outflows undergo rapid neutron capture (r-process) nucleosynthesis, making mergers promising candidates as the site of production of many heavy elements (e.g. gold, platinum, uranium, ~\citealt{korobkin:12,Wanajo2014}). Radioactive decay of the ashes of the r-process also powers bright electromagnetic counterparts to the gravitational waves emitted by mergers: the optical/infrared transients called {\it kilonovae}~\citep{Lattimer:1976,Li:1998bw,Roberts2011,Kasen:2013xka}. A kilonova was for example observed in the afterglow of the first neutron star merger detected through gravitational waves~\citep{EM170817,Cowperthwaite:2017}. 

The mass and composition of merger outflows are the main determinant of the color, duration, and luminosity of kilonovae~\citep{Barnes:2013}, while the composition of the outflows largely set the relative yields of different heavy elements as a result of r-process nucleosynthesis~\citep{korobkin:12,Wanajo2014,Lippuner2015}. Accordingly, neutrino transport is an important component in any effort to model kilonovae and r-process nucleosynthesis in mergers. Neutrino-antineutrino annihilations can also deposit a significant amount of energy in low-density regions above the remnant~\citep{Perego2017}. While, on their own, $\nu\bar\nu$ annihilations probably do not deposit enough energy to power short gamma-ray bursts (SGRBs), they may help by clearing the polar regions of baryonic matter~\citep{Fujibayashi2017}.

The main objective of a radiation transport scheme is to evolve the distribution function
$f_{(\nu)}(x^\mu,p^\mu)$ of neutrinos or photons, where $x^\mu=(t,x^i)$ are the spacetime coordinates
and $p^\mu=dx^\mu/d\lambda$ are the components of the 4-momentum of the neutrinos/photons, with $\lambda$ some affine parameter. 
Here we neglect neutrino masses, and the 4-momentum is thus
a null vector, i.e $p^\mu p_\mu=0$. The distribution function of photons and of each species of neutrinos
evolves according to Boltzmann's equation
\beq
p^\alpha \left[\frac{\partial f_{(\nu)}}{\partial x^\alpha} - \Gamma^\beta_{\alpha \gamma}
p^\gamma \frac{\partial f_{(\nu)}}{\partial p^\beta} \right] = \left[\frac{d f_{(\nu)}}{d\tau}\right]_{\rm coll}\,\,,
\eeq 
where $\Gamma^\alpha_{\beta \gamma}$ are the Christoffel symbols, and the right-hand side includes all collisional
processes. Solving Boltzmann's equation thus requires the evolution
in time of a 6-dimensional function, a very steep computational challenge. The main objective of this work is to provide a 
relatively cheap algorithm for general relativistic radiation transport which, while approximate for the numerical resolutions
that we can currently afford, asymptotes to Boltzmann's equation in the limit of infinite computational resources. While our 
algorithm can theoretically be used in various problems involving general relativistic photon or neutrino transport, we implement and test it with the problem of neutrino transport in merger simulations in mind. We implement the algorithm in the SpEC code\footnote{http://www.black-holes.org/SpEC.html}, currently used to study merging black holes and neutron stars. 

In the context of neutron star-neutron star (NSNS) and neutron star-black hole (NSBH) mergers, 
neutrinos have so far been modeled using approximate transport schemes,
with sophistication ranging from order-of-magnitude accurate leakage schemes~\citep{Deaton2013,Neilsen:2014}, to various moment schemes in which only the lowest
or two-lowest moments of $f_{(\nu)}$ in momentum space are evolved~\citep{Wanajo2014,FoucartM1:2015,Radice:2016}.
Neutrinos are also generally evolved in the gray approximation, in which information about the
energy spectrum of the neutrinos is either unavailable or, in the case of our most recent moment scheme, limited to the knowledge of the
average energy of the neutrinos~\citep{FoucartM1:2016b}. Spectral (energy-dependent) leakage and moment schemes have been
used in Newtonian simulations of post-merger accretion disks~\citep{just:2015,Perego:2016}, or in the context of core-collapse supernovae~\citep{RobertsM1:2016},
but not for NSNS/NSBH mergers. 
This is particularly problematic because neutrino cross-sections have a strong dependence
in the energy of the neutrinos.  In NSNS mergers, this leads to significant uncertainties in the computation of the composition of 
the ejecta~\citep{FoucartM1:2016a,FoucartM1:2016b}.

Moment schemes also require information about moments of $f_{(\nu)}$ that are not evolved in the simulation.
For example, in the two-moment formalism, the energy and flux density of the neutrinos are evolved, and the evolution equations
require information about the pressure tensor of the neutrinos. Semi-arbitrary choices thus have to be made in order to close the system of equations.
The most common choice in recent simulations has been the Minerbo (M1) closure~\citep{Minerbo1978}. The M1 closure, however, 
is only guaranteed to be correct in two regimes: in the optically thick limit, and for a single beam of free-streaming neutrinos. Otherwise, the closure is inaccurate. The best known
consequence of this inaccurate closure is the presence of radiation shocks whenever neutrino beams cross or converge.

To go beyond the moment formalism, at least two directions can be considered. The first is to discretize $f_{(\nu)}$ in momentum space. The cost
associated with a 6-dimensional grid is however prohibitive. Recently, a first axisymmetric simulation with a full discretization of $f_{(\nu)}$ has been
performed in the context of core-collapse supernovae~\citep{Nagakura2017}. 
While such a scheme may one day become affordable without reducing the dimensionality of the problem, it remains at this point well beyond our reach.
A second possibility is to rely on Monte-Carlo (MC) methods, randomly sampling the 6-dimensional space of $f_{(\nu)}$. In MC codes, we evolve {\it neutrino
packets} representing large numbers of neutrinos that propagate through the numerical grid. The distribution function
$f_{(\nu)}$ can be reconstructed at any point from these packets, 
within statistical errors due to the finite number of samples/packets. Recently, the MC code {\it Sedonu}
has been used to perform Newtonian evolutions of neutrinos in time-independent snapshots of supernovae simulations~\citep{Richers:2017}, while the general
relativistic code {\it bhlight} has been developed for simulations of accretion disks~\citep{Ryan2015}. While 6-dimensional general relativistic simulations 
of neutrino transport coupled to general relativistic hydrodynamics simulations of supernovae or neutron star mergers with a MC code appear closer than 
with a grid-based code, they still require a very significant investment of computational resources.

In this paper, we present the first implementation in a general relativistic hydrodynamics code of a somewhat cheaper option: a two-moment scheme 
where both the unknown moments of $f_{(\nu)}$ and any spectral information needed by the code are obtained from MC evolution of the neutrinos. Our method is developed in the spirit of the Variable Eddington Tensor algorithm implemented in the Athena code by~\cite{Davis2012} for Newtonian simulations of accretion disks, except that we use the noisier but cheaper Monte-Carlo methods to provide the closure instead of the short-characteristic method used by~\cite{Davis2012} 

The fact that we have to evolve neutrinos
with a MC scheme may give the impression that our algorithm does not provide any advantage over a pure MC evolution. However, using MC 
only as a way to close the equations for the evolution of the moments allows us to take two important, cost-saving shortcuts. First, we use MC to compute time-averaged moments of $f_{(\nu)}$.
At fixed number of neutrino packets on the grid, this allows us to reduce the statistical error in the computation of the moments without increasing the
cost of the simulation. This comes, however, at the cost of smoothing over time the value of these moments. Second, we can completely side-step a common issue in MC simulations: the high cost of evolving optically thick regions, where many packets are constantly created and reabsorbed. MC codes can partially avoid this issue by 
switching to a different treatment of the neutrinos in optically thick regions, e.g. by using MC methods to approximate a diffusion equation through optically thick regions. We instead simply rely on the moment scheme in optically thick regions, where it is very accurate, and only evolve MC packets below a certain optical depth. 

The use of a two-moment scheme offers another practical advantage: it is generally much simpler to satisfy the Hamiltonian and Momentum constraints of Einstein's equations, as well as conservation laws, in a grid-based moment scheme than in a pure MC evolution.
Finally, and most importantly, as the MC evolution asymptotes to the exact solution to Boltzmann's equation, and as the evolution of the moments can take from that MC evolution all of the missing information about higher moments of the distribution function and about neutrino spectra that would otherwise be approximated using analytical prescriptions, the algorithm satisfies our requirement that it theoretically converges to a true solution of the transport problem as more computational resources become available. 

At this point, however, we should mention an important caveat: while the test results obtained with our code, and described in this manuscript, are so far encouraging, we have not rigorously proven that the coupled MC-Moment system is numerically stable. Whether any additional work is required to guarantee that the coupled MC-moment equations are well-behaved for realistic astrophysical simulations is an important open questions, that will require additional investigation. In practice, in the near future, we plan to use our algorithm in two distinct ways. First, we will use the MC evolution to compute the neutrino spectra, the absorption and scattering opacities, and the rate of $\nu \bar\nu$ annihilation in low-density regions above a neutron star merger remnant, and to obtain detailed information about the distribution function of neutrinos in neutron star mergers. This provides a significant improvement over a gray two-moment scheme, but not the desired outcome of an algorithm converging to a true solution to Boltzmann's equation (as, e.g., the neutrino pressure tensor would still be approximated using the M1 closure). In a second stage, we will use the fully coupled system presented in this manuscript, possibly improved to handle any additional stability issues encountered in realistic astrophysical systems. We note that our approach to this problem makes it easy to rely on either the MC evolution or analytical approximation to provide any of the missing information in the evolution of the moments of $f_{(\nu)}$. This makes switching from a ``partially coupled'' MC-Moment scheme to a ``fully coupled'' MC-Moment scheme, as appropriate for any given project, a fairly easy task.

In the rest of this paper, we assume that $G=c=1$. For subscripts and superscripts, greek letters are spacetime indices going from 0 to 3, 
and roman letters are spatial indices going from 1 to 3. When a black hole is involved in the simulation its mass is $M_{\rm BH}=1$. Otherwise, we work
with an arbitrary mass unit $M=1$ (code units), or with $M_\odot=1$. A table summarizing the meaning of the various symbols used throughout this work 
is provided at the end of the manuscript (Table~\ref{tab:notations}). 
We begin with a description of our numerical method (Sec.~\ref{sec:methods}), then demonstrate the performance of our algorithm 
in a few idealized test cases (Sec.~\ref{sec:tests}), and finally conclude with a discussion of the strengths and limitations of our algorithm in its current form, and of
potential applications (Sec.~\ref{sec:conclu}).

\section{Numerical Methods}
\label{sec:methods}

The SpEC code can be used to evolve Einstein's equations of general relativity coupled to the general
relativistic equations of hydrodynamics. In this paper, however, we focus on the development of a new neutrino
transport scheme for SpEC. For the code tests presented here, we do not evolve Einstein's equations, while the fluid
is at most evolved through its coupling to the neutrinos. From our experience developing a two-moment neutrino transport
scheme in SpEC, we do not expect the full coupling to Einstein's equations and the equations of hydrodynamics to create
 new problems in our evolutions, although whether this remains true for the coupled MC-moment scheme developed here
still has to be tested in practice. In the following sections, we first define the reference frames in which we solve the 
radiation transport problem (Sec.~\ref{sec:def}), then discuss the implementation of the two-moment (Sec.~\ref{sec:M1})
and Monte-Carlo (Sec.~\ref{sec:MC}) transport schemes, the methods used to couple the two schemes (Sec.~\ref{sec:MCM1Coupling}),
and finally how the various pieces of our radiation transport algorithm fit within the evolution of the general relativistic
radiation hydrodynamics equations in SpEC (Sec.~\ref{sec:fullCoupling}).

\subsection{Definitions and reference frames}
\label{sec:def}

We assume a 3+1 decomposition of spacetime. The 3+1 decomposition relies on a foliation of spacetime
into slices of constant time coordinate $t$, with timelike unit normal $n^\mu = \alpha^{-1}(t^\mu - \beta^\mu)$ and line element
\beqn
ds^2 &=& g_{\mu\nu} dx^\mu dx^\nu \nonumber\\
	&=& -\alpha^2 dt^2 + \gamma_{ij} (dx^i + \beta^i dt)(dx^j + \beta^j dt),
\eeqn
where $g_{\mu\nu}$ is the 4-metric, $\gamma_{ij}$ the 3-metric on a slice of constant $t$, $\alpha$ the lapse, 
and $\beta^i$ the shift vector. Our numerical
grid is discretized in the spatial coordinates $x^i$. We will refer to the coordinates $(t,x^i)$ as the {\it grid frame}.

The fluid is described by its baryon density $\rho$, temperature $T$, electron fraction $Y_e$, and 4-velocity $u^\mu$.
Two special observers will play an important role in the description of our neutrino transport algorithm: {\it inertial observers},
whose timeline is tangent to $n^\mu$, and {\it comoving observers}, whose timeline is tangent to $u^\mu$. We also
define the coordinates of the {\it fluid rest frame} $(t',x^{i'})$, which are defined at a point so that 
\beq
ds^2 = \eta_{\mu'\nu'} dx^{\mu'} dx^{\nu'},
\eeq
with $\eta_{\mu\nu}$ the Minkowski metric, and $(t')^\mu = u^\mu$. We construct these local coordinates from an orthonormal tetrad 
${\hat{e}^{(\lambda')}_\mu}$, with $g^{\mu\nu} \hat{e}^{(\lambda')}_\mu \hat{e}^{(\kappa')}_\nu=\eta^{\lambda'\kappa'}$, and $\hat e_{(t')}^\mu = u^\mu$.
The three other components of the tetrad are obtained by applying Gramm-Schmidt's algorithm to the three vectors $V_{(i)}^\mu = \delta^\mu_i$ ($i=1,2,3$).
The orthonormal tetrad 
${\hat{e}_{(\lambda')}^\mu}$ and the corresponding one-forms ${\hat{e}^{(\lambda')}_\mu}$ are precomputed and stored for each grid cell, and can be used to easily perform transformations from the fluid rest frame coordinates to grid coordinates (and vice-versa) by simple matrix-vector multiplications. 
We note that their are many alternative methods to choose
a convenient tetrad for radiation transport, e.g.~\cite{Lindquist:1966,Caradall:2013,Shibata:2014,Nagakura:2017b}.
Because we do not assume any symmetry when constructing our tetrad, there is no obvious geometrical meaning to this tetrad, besides the fact that its first element is tangent to
the world line of an observer comoving with the fluid. In special relativistic problems and for a fluid velocity aligned with a coordinate axis, our coordinate transformation from the
grid frame to the fluid rest frame is a standard Lorentz boost. For fluid velocities of arbitrary orientation, it already differs from the standard choice by a spatial rotation in the fluid rest 
frame.

Unless otherwise specified, vector and tensor components are expressed in the grid coordinates. The fluid rest frame is mostly used to compute 
neutrino-matter interactions. Quantities expressed in the fluid rest frame will be primed, e.g. $p^{\mu'}$ for the momentum of the neutrinos in the fluid rest frame.

\subsection{Two-moment transport}
\label{sec:M1}

The first layer of our transport scheme is based on the two-moment formalism. A two-moment scheme was already implemented
in the SpEC code, and has been used for the study of NSBH~\citep{FoucartM1:2015} and NSNS~\citep{FoucartM1:2016a,FoucartM1:2016b} mergers. The general idea of the scheme, proposed by~\cite{Thorne:1981}
and~\cite{shibata:11}, is to evolve moments of
the distribution function $f_{(\nu)}$. These can be defined from the stress-energy tensor of the neutrinos,
\beq
T^{\mu\nu}_{(\nu)} = E n^\mu n^\nu + F^\mu n^\nu + n^\mu F^\nu + P^{\mu\nu}.
\eeq
Here, $E$ is the energy density of the neutrinos measured by an inertial observer. The energy flux $F^\mu$ and pressure tensor
$P^{\mu\nu}$ are both normal to $n^\mu$, i.e. $F^\mu n_\mu = P^{\mu \nu} n_\mu = P^{\mu\nu}n_\nu = 0$. Evolution equations for $\tilde E = \sqrt{\gamma}E$,
$\tilde F_i = \sqrt{\gamma} F_i$, with $\gamma$ the determinant of the 3-metric $\gamma_{ij}$, are obtained by integrating Boltzmann's equation:
\beqn
\partial_t \tilde E &+& \partial_j(\alpha \tilde F^j -\beta^j \tilde E)\label{eq:Enu}\\
&=&\alpha (\tilde P^{ij}K_{ij} -\tilde F^j \partial_j \ln{\alpha} - \tilde S^\alpha n_\alpha)\nonumber\,\,,\\
\partial_t \tilde F_i &+& \partial_j(\alpha \tilde P^j_{i} -\beta^j \tilde F_i) \label{eq:Fnu}\\
&=&(-\tilde E\partial_i\alpha+\tilde F_k\partial_i\beta^k+\frac{\alpha}{2} \tilde P^{jk}\partial_i \gamma_{jk}+\alpha \tilde S^\alpha \gamma_{i\alpha})\nonumber\,\,,
\eeqn
where $K_{ij}$ is the extrinsic curvature of the current spatial slice.
To close this system of equations, we require two pieces of information: the pressure $\tilde P_{ij} = \sqrt{\gamma} P_{ij}$ and the source terms 
$\tilde S^\alpha = \sqrt{\gamma} S^\alpha$. We define
\beq
P_{ij} = \pi_{ij} E,
\eeq
$\pi_{ij}$ being the Eddington tensor. In this manuscript, we also write the source terms
\beq
S^\alpha = \eta u^\alpha - \kappa_a J u^\alpha - (\kappa_a + \kappa_s) H^\alpha.
\eeq
Here, $\kappa_a$ is the absorption opacity, $\kappa_s$ the scattering opacity, and $\eta$ the emissivity.
The Eddington tensor and opacities are quantities to be provided 
by the MC code, while the emissivity $\eta$ is taken from tabulated values for neutrino-matter interactions (see Secs.~\ref{sec:kappa} and \ref{sec:emission}). 
We note that this form for $S^\alpha$, chosen for its convenience in the tests used in this manuscript, 
excludes potentially important processes, e.g. pair annihilations (which couple neutrinos and anti-neutrinos) or
inelastic scatterings. Energy and momentum deposition from these processes can however be computed from the MC evolution, 
and added as source terms to the evolution of the moments as needed. It is fairly easy to do so as long as the additional
source terms are not too large, and can thus be treated explicitly
 
The moments $J$ and $H^\alpha$ are, respectively,
the energy density and flux density measured by an observer comoving with the fluid (but with vector components in grid coordinates).
They can be defined from the stress-energy tensor of the neutrinos
\beq
T^{\mu\nu}_{(\nu)} = J u^\mu u^\nu + H^\mu u^\nu + u^\mu H^\nu + S^{\mu\nu},
\eeq
with $S^{\mu\nu}$ the pressure tensor measured by a comoving observer, and $H^\mu u_\mu = S^{\mu\nu}u_\mu = S^{\mu\nu}u_\nu=0$. 
$J$ and $H^\mu$ can thus be computed as functions of $E$, $F_i$, and $\pi_{ij}$,
by taking projections of $T^{\mu\nu}_{(\nu)}$. 

An equivalent, and more intuitive definition of the moments can be obtained if we start from the comoving moments
\beqn
J &=& \int d\nu \nu^3 \int d\Omega f_{(\nu)}\\
H^\mu &=& \int d\nu \nu^3 \int d\Omega f_{(\nu)} l^\mu\\
S^{\mu\nu} &=& \int d\nu \nu^3 \int d\Omega f_{(\nu)} l^\mu l^\nu
\eeqn
with $\nu$ the neutrino energy in the fluid rest frame, $\int d\Omega$ integrals over solid angle on a unit sphere in momentum space, and
\beq
p^\mu = \nu (u^\mu + l^\mu),
\eeq
the 4-momentum of neutrinos, with $l^\mu u_\mu=0$.

We use high-order finite volume methods to evolve $(\tilde E,\tilde F_i)$. A locally implicit time stepping allows us to handle stiff neutrino-matter interaction terms, while the flux terms are computed explicitly. The only terms to be treated implicitly in the equations are those containing the source terms $S^\alpha$. Thanks to the linearity of the equations in $(\tilde E,\tilde F_i)$, we can solve the implicit problem exactly by inverting a  4x4 matrix for each neutrino species at each point. 

In practice, we compute the fluxes $(\alpha \tilde F^j -\beta^j \tilde E)$ and $(\alpha \tilde P^j_{i} -\beta^j \tilde F_i)$ on cell faces
using the HLL approximate Riemann solver~\citep{HLL}, and a fifth-order WENO reconstruction of $E$, $F_i/E$, $\pi_{ij}$ on cell faces from their cell-centered values~\citep{Liu1994200,Jiang1996202}. 
The time discretization uses an 
implicit second-order Runge-Kutta method: to evolve the system by a time step $\Delta t$, we first take a test step $\Delta t/2$, with the fluxes and explicit source terms 
computed at the beginning of the time step. We then take a full step $\Delta t$, with the fluxes and explicit source 
terms evaluated from the result of the half step. 

To compute the HLL fluxes, we need the characteristic speeds of the system. For the linear system considered here, the minimum and maximum speeds across a face in the direction $d$ are
\beq
c_\pm = -\beta^d \pm \alpha \sqrt{\pi^{dd}}.
\eeq

Finally, we note that the two-moment scheme is corrected in very optically thick cells (i.e. cells in which $[\kappa_a + \kappa_s] \Delta x \gtrsim 1$, with $\Delta x$ the grid 
spacing). Without such a correction, the diffusion of neutrinos in optically thick regions is set by numerical viscosity. This correction effectively transitions between a two-moment scheme and a one-moment scheme, with $F_i(E)$ in optically thick regions set by its known value in the diffusion limit. In practice, we follow the method of~\cite{Jin1996}. Details of its implementation in the SpEC code are discussed in~\cite{FoucartM1:2015}. 

\subsection{MC transport}
\label{sec:MC}

The second and most expensive layer of our transport scheme is a MC algorithm, which we implement in SpEC for this work. Our MC methods are largely inspired from
the general relativistic radiation hydrodynamics code bhlight~\citep{Ryan2015}. We refer the reader to that work for more details on the derivation of a general relativistic MC scheme, and focus here on a summary of the algorithm, as well as on the additional work required to use a MC algorithm to close the two-moment equations.

In a MC scheme, we evolve {\it neutrino packets} that sample the distribution function of neutrinos, $f_{(\nu)}$. The ensemble of $P$ packets in the simulation at time $t$, with packet $k$ representing $N_k$ neutrinos at the spatial coordinates $x^i_k$ and with 4-momentum $p^\mu_k$, provides an estimate of the distribution function
\beq
f_{(\nu)} \sim f_{(\nu),MC} = \sum_{k=1}^{P} N_k \delta^3(x^i-x^i_k) \delta^3(p_i-p_i^k).
\eeq
We note that the use of lower indices in $p_i$ and upper indices in $x^i$ is required for this equation
to be a relativistic invariant. To couple the MC evolution to the moment formalism, we also rely on the MC estimate of the neutrino stress-tensor
\beq
T^{\mu\nu}_{(\nu),MC} =  \sum_{k=1}^{P} N_k \frac{p_k^\mu p_k^\nu}{\sqrt{\gamma}\alpha p_k^t}  \delta^3(x^i-x^i_k).
\eeq

In a MC transport scheme, Boltzmann's equation for $f_{(\nu)}$ can be translated into prescriptions for the creation, annihilation, scattering and propagation of the neutrino packets sampling $f_{(\nu)}$, which we discuss in the following sections. In SpEC, the state of a neutrino packet is entirely determined by its grid coordinates $(t,x^i)$, momentum $p_i$, and neutrino species, as well as the number of neutrinos $N_k$ represented by the packet. 
As we neglect the masses of neutrinos, the fourth component of the momentum can easily be computed from $p_i$, e.g. $p^t = \sqrt{\gamma^{ij} p_i p_j} /\alpha$. The spatial components of the grid coordinates $x^i$ and momenta $p_i$ are evolved in time, while $N_k$ and the neutrino species remain constant during the evolution of a packet. We also regularly need the fluid rest frame energy of the neutrinos in the packet,
\beq
\nu = \alpha W p^t - \gamma^{ij} u_i p_j,
\eeq
where $W=\sqrt{1+\gamma^{ij} u_i u_j}$ is the Lorentz factor of the fluid with respect to an inertial observer.

For high-accuracy evolution of neutrino packets in the MC framework, high-order interpolation of the fluid and metric variables to the position of a neutrino packet would be desirable. However, this would significantly increase the cost of the MC scheme. As limiting the cost of the MC algorithm is our main concern at this point, we make a simpler, less accurate choice: we use cell-centered values of the fluid variables, metric variables, and derivatives of the metric, which are already computed during the evolution of the general relativistic equations of hydrodynamics. We will see in Sec.~\ref{sec:tests} that this is unlikely to be an important contribution to the error budget of our simulations.

\subsubsection{Tabulated neutrino-matter interactions}
\label{sec:kappa}

In this manuscript, we ignore inelastic scatterings and reactions involving two or more neutrinos. Neutrino-matter interactions are described by a neutrino emissivity $\eta(\rho,T,Y_e,E_{(\nu)})$, absorption opacity $\kappa_a(\rho,T,Y_e,E_{(\nu)})$, and elastic scattering opacity $\kappa_s(\rho,T,Y_e,E_{(\nu)})$, for each species of neutrinos and antineutrinos. The exact interpretation of these variables in the context of a MC algorithm will be discussed in more detail in the following sections.  

We use tabulated values of these quantities produced by the open-source NuLib library~\citep{OConnor2010}, which provides a flexible framework to include a range of neutrino-matter interactions of importance to the merger problem, and let us choose the density of the table in the fluid variables $(\rho,T,Y_e)$. We use linear interpolation in $[\log{(\rho)}, \log{(T)}, Y_e]$ to interpolate between tabulated values. We discretize the neutrino spectrum into $N_E$ energy bins, with bounds $(E_{0}=0,E_{1},...,E_{N_E})$.
NuLib provides us with values of the opacities at the center of each energy bin. We obtain the opacities at other energies by interpolating linearly in $\log{(E_{(\nu)})}$. 
NuLib also provides us with the emissivity per unit volume, integrated over each energy bin. The tabulated values guarantee that $\eta$ and $\kappa_a$ satisfy Kirchoff's Law, i.e. that for each energy bin $\eta/\kappa_a$ is the energy density of neutrinos in equilibrium with the fluid, integrated over that bin. By enforcing Kirchoff's law in that manner, we make sure that the neutrinos reach the desired equilibrium distribution in optically thick regions.
For testing purposes, we also implement an alternative framework in which $\eta, \kappa_a, \kappa_s$ are provided as functions of the spacetime coordinates.

We note that incorporating inelastic scattering of neutrinos by nucleons, electrons, and nuclei into the MC evolution is not particularly complex from an algorithmic point of view, but requires very large tables for the inelastic scattering cross-sections, coupling all energy bins. Neutrino-antineutrino annihilations into $e^+ e^-$ pairs are more challenging if one wants to account 
for the blocking factor of the electrons. Ignoring that factor (which is probably acceptable in the low-density regions where pair annihilation may play an important role), one can define the energy deposition due to pair annihilation from the moments $(E,F_i,P_{ij})$ of the distribution function of each neutrino species~\citep{Fujibayashi2017}, which we already have at our disposal. We plan to incorporate $\nu \bar\nu$ annihilations in that approximation, once we begin to use our code to study compact binary mergers.

\subsubsection{Emission}
\label{sec:emission}

To sample the emission of neutrinos, we compute for each cell of volume $\Delta V=\Delta x^1 \Delta x^2 \Delta x^3$ and time interval $\Delta t$ the number of neutrinos packets $N_p$ emitted with a fluid rest frame energy within a given energy bin $[E_{b-1},E_b]$. Taking advantage of the invariance of the 4-volume $(\sqrt{-g} \Delta V \Delta t)$, and defining $\eta_b(\rho,T,Y_e)$ as the tabulated emissivity per unit volume in that energy bin, we find
\beq
N_p \approx \sqrt{-g} \Delta V \Delta t \frac{\eta_b(\rho,T,Y_e)}{E_p},
\eeq
where $E_p$ is the desired energy of each neutrino packet in the fluid frame and $\sqrt{-g}=\alpha \sqrt{\gamma}$. 
$E_p$ is provided as an input to the code, and effectively sets the number of packets evolved by the MC algorithm. If $[N_p] $ is the largest integer smaller than $N_p$, then we create $[N_p]$ neutrinos packets, with a probability $(N_p-[N_p])$ of creating one additional packet. 

All packets are created with a fluid rest frame energy $\nu=0.5*(E_{b-1}+E_b)$, and represent a number of neutrinos $N_k = E_p/\nu$. The choice to initialize all neutrinos with the energy of the center of the bin was proposed by~\cite{Richers:2017} and has the advantage to be consistent with the way Kirchoff's law is enforced in the NuLib tables (i.e. by balancing the emissivity $\eta_b$ integrated over the entire energy bin with the opacity $\kappa_a(\nu)$ at the center of the bin). The time of creation and initial position of the packets are drawn from uniform distributions in $(t,x^i)$ within the volume $\Delta V$ and time interval $\Delta t$. We note that the choice of uniform sampling in the coordinates $x^i$ could be a significant approximation when using curvilinear coordinates, or if the metric varies significantly over the length of a cell. It may be necessary to use sampling methods which take into account variations in the proper volume element within a cell for applications requiring higher accuracy and/or for non-cartesian grid structures. 

The initial direction of propagation of the neutrino packets is drawn so that the packets are isotropically distributed in the fluid rest frame. The 4-momentum of a neutrino in a given packet is, in that frame and using the orthonormal tetrad defined in Sec.~\ref{sec:def},
\beq
p^{\mu'} = \nu \left(1,\sin\theta \cos\phi,\sin\theta \sin\phi,\cos\theta \right).
\eeq
We draw $\cos(\theta)$ from a uniform distribution in the range $[-1,1]$ and $\phi$ from a uniform distribution in the range $[0,2\pi]$. The 4-momentum of neutrinos in grid coordinates can then be computed using that same orthonormal tetrad:
\beqn
p^t &=& {\hat{e}_{(\lambda')}^t} p^{\lambda'},\\
p_i &=& g_{i\mu} {\hat{e}_{(\lambda')}^\mu} p^{\lambda'} =\delta_{\kappa' \lambda'} {\hat{e}^{(\kappa')}_i} p^{\lambda'}.
\eeqn
The angles $\theta$, $\phi$ used in this sampling process are convenient to obtain an isotropic distribution in the fluid rest frame. As opposed to angles defined with respect to the radial direction of a spherical grid, however, they have no clean interpretation. The ``polar'' angle $\theta$ is the angle between the direction of propagation of the neutrinos and the last element of the orthonormal tetrad constructed by applying Gramm-Schmidt's algorithm to the coordinate axes of of the grid on which we evolve the moment's equations. In curved spacetime, this is not a particularly meaningful vector.

We note that, after that transformation and summing over all bins, the total energy $\Delta E$ of the neutrinos created within a cell is, on average and as measured by an inertial observer, $\Delta E=\Delta V \Delta t \sqrt{\gamma} u^t (\sum_b \eta_b)$. By comparing this result with the definition of the energy integrated emissivity $\eta$ used in the evolution of the moments equations, we find that, unsurprisingly, $\eta = \sum_b \eta_b$. The emissivity in the gray two-moment transport can thus be obtained directly from the tabulated emissivities used by the MC algorithm.

\subsubsection{Initialization of a cell}

The MC algorithm also needs a prescription for the initialization of packets within a cell at the beginning of a time step. Such initialization is required at the beginning of the first time step, as well as in some optically thick regions discussed in Sec.~\ref{sec:largeKa}. In optically thick regions, we want these packets to sample the equilibrium distribution of neutrinos. In optically thin regions, we do not have any particularly good guess to rely on, and thus do not create any neutrino packets at the initial time. We assume that the number density of neutrinos, on the initial slice and within the energy bin $[E_{b-1},E_b]$ is, in the fluid rest frame,
\beq
dn = d^3x^{i'} d\Omega \frac{\eta_b}{4\pi \nu[\kappa_a(\rho,T,Y_e,\nu)+a]},
\eeq
with $\nu=0.5*(E_{b-1}+E_b)$.
The constant $a$ is semi-arbitrary, and sets the initial energy density of neutrinos in regions of moderate optical depth. At $t=0$ and for the problems presented here, we choose $a=(\kappa_a+\kappa_s)^{-1} L^{-2}$, where $L$ is a length scale comparable to the typical length scale for variations of $\kappa_{a}$. In our tests, we use $L=GM/c^2$, with $M=1$ (in code units) for idealized tests and $M=M_\odot$ for our test using a core-collapse fluid profile.

A slight complication when sampling the neutrino distribution function on a spatial slice, also discussed in~\cite{Ryan2015}, is that the 3-volume $\sqrt{\gamma} \Delta V$ is not a relativistic invariant. Accordingly, one has to be careful when sampling in the inertial frame a distribution function which is known only in the fluid rest frame. We can rely on the invariance of $\sqrt{-g} \Delta V p^t$ and the fact that $\sqrt{-g}=1$ and $p^{t'}=\nu$ in the fluid rest frame to derive the ratio between the volume element in the fluid rest frame, $\Delta V' = \Delta x^{1'} \Delta x^{2'} \Delta x^{3'}$, and the volume element in the grid frame, $\Delta V$: $\Delta V'/\Delta V = \sqrt{-g} p^t/\nu$. Using this result, we sample the distribution of neutrinos by creating $N^0_p$ neutrino packets
\beq
N^0_p =  \sqrt{-g} \Delta V \frac{\eta_b(\rho,T,Y_e)}{E_p [\kappa_a(\rho,T,Y_e,\nu)+a]}.
\eeq
Each packet represents neutrinos with fluid rest frame energy $\nu$. 
The non-integer part of $N^0_p$ is treated as a probability to emit one more packet, the initial position of each packet is drawn from a uniform distribution in $\Delta V$, and the momentum of the neutrinos is drawn from an isotropic distribution in the fluid frame, as for the main emission procedure. Each packet represents $N_k = (E_p/\nu) \times (p^t/\nu)$ neutrinos, and not $E_p/\nu$ neutrinos, to account for the ratio $\Delta V/\Delta V'$. 

\subsubsection{Propagation, absorption, and scattering}
\label{sec:propagation}

In our code, the evolution of MC packets currently involves three types of operation: the propagation of packets along null geodesics, as well as absorption and scattering of packets sampling neutrino-matter interactions. To evolve a neutrino packet by a time interval $\Delta t$, we first determine whether the packet is free-streaming, or whether it is absorbed or scattered by the fluid. Absorption and scattering probabilities can be computed from the infinitesimal optical depth along a geodesic, $d\tau = \kappa \nu d\lambda =(\kappa \nu/p^t) dt$, with $d\lambda$ the increment in the affine parameter ($p^\mu = dx^\mu/d\lambda$). The time interval before the first absorption/scattering is then 
\beq
\Delta t_{s,a} = - \log{(r_{s,a})} \frac{p^t}{\kappa_{s,a}\nu},
\eeq
with $r_{s,a}$ drawn from a uniform distribution in $(0,1]$. We then determine the smallest of the three time intervals $(\Delta t,\Delta t_a,\Delta t_s)$. If $\Delta t$ is the smallest time interval, the packet is propagated by $\Delta t$ without interacting with the fluid. If $\Delta t_a$ is the smallest interval, the packet is propagated by $\Delta t_a$ and then absorbed (i.e. removed from the simulation). Finally, if $\Delta t_s$ is the smallest interval, the packet is propagated by $\Delta t_s$, then scattered by the fluid. After scattering we begin a new time step with $\Delta t \rightarrow \Delta t - \Delta t_s$. Scattering is performed in the fluid rest frame. As we only consider isotropic elastic scattering, we simply redraw the 4-momentum $p^{\mu'}$ (at constant fluid rest frame energy $\nu$), from the same istropic distribution as during packet creation.

Propagation of a packet along null geodesics is performed following the prescription of~\cite{Hughes1994},
\beqn
\frac{dx^i}{dt} &=& \gamma^{ij} \frac{p_j}{p^t} - \beta^i,\\
\frac{dp_i}{dt} &=& -\alpha (\partial_i \alpha) p^t + (\partial_i \beta^k) p_k - \frac{1}{2} (\partial_i \gamma^{jk}) \frac{p_j p_k}{p^t}. \label{eq:EvP}
\eeqn
We use the same second-order Runge-Kutta time stepping as for the two-moment algorithm, with the caveat that we use metric quantities evaluated at the cell center closest to the initial position of the packet. We only switch the cell from which we gather the fluid and metric variables at the end of a time step. This naturally imposes a limit on the timestep $\Delta t \lesssim A \Delta x^i/c_{\rm max}$, where $A$ reflects our tolerance for how far packets potentially move into a neighboring cell before we use updated values of the metric, and $c_{\rm max}$ is the maximum grid-coordinate value of the speed of light. In most tests, we choose $A\sim 1/3$, which is comparable to the Courant condition for the evolution of the fluid and of the moments $(E,F_i)$.

\subsubsection{Optically thick regions}
\label{sec:largeKa}

In theory, we now have at our disposal all of the pieces required for a basic MC scheme: initialization of the simulation, emission of neutrino packets, and interactions with the fluid. We could move on to the computation of the backreaction on the evolution of the moments. However, this would be prohibitively expensive for at least two reasons. First, the scheme would continuously create and absorb a large number of packets in optically thick regions, where a very small fraction of the emitted packets survive any single time step. Second, in regions in which the scattering opacity $\kappa_s \Delta t \gg 1$, the code would spend too much time propagating neutrino packets over a large number of time intervals $\Delta t_s \ll \Delta t$. Both of these issues can, however, be handled through much cheaper methods.

To do so, we define a 4-dimensional grid in $(x^i,\nu)$, with the spatial discretization being provided by the grid on which we evolve the moments, and the energy discretization provided by the binning used when generating the NuLib table for neutrino-matter interactions. We define three types of cells on this grid. Cells with 
$\sqrt{\kappa_a (\kappa_a+\kappa_s)} \geq \kappa_{\rm crit}$ are {\it optically thick cells}. There, we assume that the neutrinos are in thermal equilibrium with the fluid. Cells which are not optically thick and satisfy $\kappa_s \Delta t' \geq t_{\rm diff}$ (with $\Delta t' = \Delta t/u^t$ the time step in the fluid rest frame) are called {\it high-scattering cells}. There, we assume that the diffusion equation is a good approximation to the evolution of the energy density of neutrinos. Finally, cells which are neither optically thick nor high-scattering use the standard MC algorithm outlined in the previous sections. The parameters $\kappa_{\rm crit}$ and $t_{\rm diff}$ can be specified at run time. It is important to note that for our algorithm to formally converge to a solution of Boltzmann's equations, we should take $\kappa_{\rm crit} \rightarrow \infty$ and $t_{\rm diff} \rightarrow \infty$, in addition to increasing the number of neutrino packets and the accuracy of the evolution of the moments. For the grid spacings and number of packets that we can currently afford, however, the errors introduced by the use of approximate methods in optically thick / high-scattering cells are subdominant.

As our algorithm couples the MC evolution to the evolution of the moments, we have a fairly simple solution to the costly evolution of optically thick cells. We simply skip the MC algorithm in optically thick regions, and rely solely on the two-moment scheme in that regime. The two-moment scheme is known to perform well in the diffusion regime, as long as some corrections are applied to the fluxes in order to limit the impact of numerical dissipation. A cell is {\it masked} if $\sqrt{\kappa_a (\kappa_a+\kappa_s)} \geq \kappa_{\rm crit}$. Any neutrino packet which ends a time step in a masked cell is removed from the simulation. When coupling the MC and moment schemes, we assume that neutrinos are in thermal equilibrium with the fluid in masked cells. The proper choice for $\kappa_{\rm crit}$ depends on the typical length scale over which $\kappa_{a,s}$ vary. We note that we only assume a thermal distribution of neutrinos for the computation of high-order moments and of $\kappa_{a,s}$. The evolution of the moments computes the diffusion of neutrinos through optically thick regions without requiring exact thermal equilibrium.

Entirely neglecting masked cells may create errors close to the boundary of the masked region. To provide an effective boundary condition to the MC algorithm, we evolve any masked cell that shares some boundary with an evolved cell (in 4-dimensional space ($x^i,\nu$), and including 1D, 2D, and 3D interfaces), using a modified MC algorithm. All packets that finish a time step in a masked boundary cell are removed from the simulation. Packets sampling an equilibrium distribution of neutrinos are then redrawn at the beginning of each time step, as during the initialization of the MC scheme (but setting the constant $a=0$). The emission of neutrino packets during a time step proceeds as in evolved cells. However, we modify $E_p$ so that no more than $N_{\rm max}$ packets are created in a single energy bin of a single boundary cell. In this paper, we use $N_{\rm max}=300$. While destroying and re-creating packets in these cells at each time step may seem highly inefficient, we note that this generally happens in cells where the number of packets required to sample the equilibrium distribution of neutrinos is comparable to the number of packets created during a time step. The additional use of a strict limit $N_{\rm max}$ on the number of packets created within each of these cells guarantees that our effective boundary conditions does not become an important cost in the MC algorithm.

\subsubsection{Approximate treatment of high-scattering regions}
\label{sec:highKs}

\begin{figure}
\includegraphics*[width=0.49\textwidth]{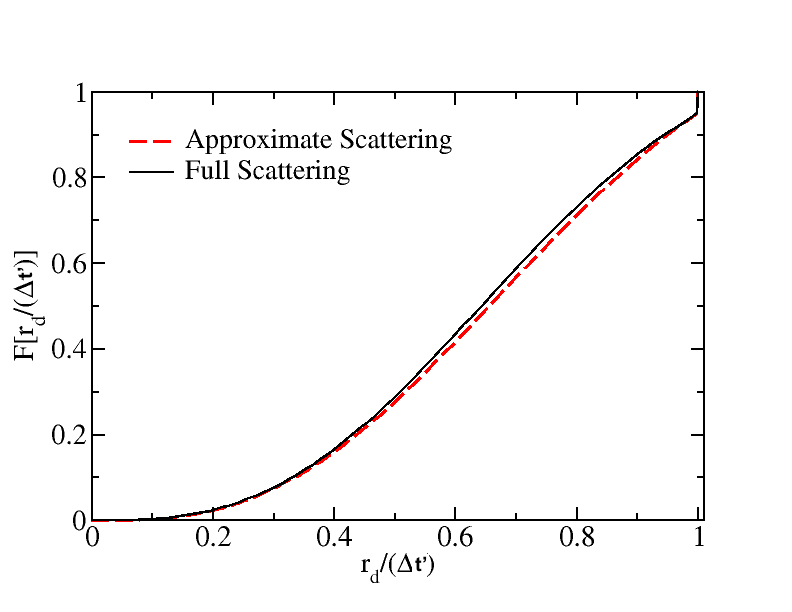}
\caption{Distribution function of $r_d/(\Delta t')$, the ratio of the distance covered by a packet to the elapsed proper time, for $\kappa_s \Delta t' =3$.
We show the approximate distribution function used in our code, and the result of scattering experiments where $10^8$ MC packets are propagated with scatterings
properly treated as individual events. Even for $\kappa_s \Delta t' =3$, the agreement between the exact and approximate treatment is quite good (and it only improves as $\kappa_s$ increases).}
\label{fig:ScatDis}
\end{figure}

In high scattering regions, instead of treating each scattering event individually, we rely on the expected diffusion velocity of packets away from their original location in the fluid rest frame. Solving the diffusion equation in spherical symmetry, we get that the probability distribution for a packet to move a distance $r_d$ from its original location in the fluid 
rest frame, over a time interval $\Delta t'$ in that frame, is
\beqn
f(\tilde r) &=& \frac{4}{\sqrt{\pi}} \tilde r^2 \exp{(-\tilde r^2)},\\
\tilde r &=& \sqrt{\frac{3\kappa_s\Delta t'}{4}} r_d.
\eeqn
We could tabulate the function $\tilde r(P)$, defined implicitly by 
\beq
\label{eq:P}
\int_0^{\tilde r(P)} f(\tilde r) d\tilde r = P,
\eeq
and then randomly draw $P$ from a uniform distribution in $[0,1]$ to obtain $r_d(P)$. In fact, we find that we can obtain a better approximation of the diffusion limit for moderate values of $\kappa_s \Delta t'$ if we set
\beq
\label{eq:Pnorm}
\int_0^{\tilde r(P)} f(\tilde r) d\tilde r = P \frac{ \int_0^{\sqrt{3\kappa_s\Delta t'/4}\Delta t'} f(\tilde r) d\tilde r}{1-\exp{(-\kappa_s\Delta t')}},
\eeq
and set $r_d=\Delta t'$ whenever that formula predicts $r_d>\Delta t'$, to avoid violating causality. Cases in which $r_d >\Delta t'$ are interpreted as packets that do not experience any scattering events, and thus behave as free-streaming neutrinos. 
Eq.~\ref{eq:Pnorm} is obtained by renormalizing Eq.~\ref{eq:P}, using the constraint that the probability for a packet to avoid any scattering (i.e. to get $r_d>\Delta t'$) is $\exp{(-\kappa_s\Delta t')}$. With that normalization, we find good agreement between the approximate distribution of $r_d$ and the distribution obtained by treating scattering events individually, for optical depths as low as $\kappa_s \Delta t' \sim 3$ (see Fig.~\ref{fig:ScatDis}).

In practice, when evolving a neutrino packet, we always evolve the packet exactly up to its first interaction. If that interaction is a scattering event, and the remaining time in the time step is such that $\kappa_s \Delta t' > t_{\rm diff}$, we continue the time step using the diffusion approximation. We then draw a new absorption time $\Delta t_a$ and evolve the packet by $\Delta t = \min(\Delta t,\Delta t_a)$ as follows. We randomly draw $P$ from a uniform distribution in $[0,1]$ to obtain $f_{\rm free}=r_d/\Delta t'$. We interpret $f_{\rm free}$ as follow: we transport the neutrino packet with the fluid for a time $\Delta t'_{\rm fl}=(1-f_{\rm free}) \Delta t'$, and then propagate the packet for a time $\Delta t'_{\rm free} = f_{\rm free} \Delta t'$ in a random direction drawn from an isotropic distribution in the fluid rest frame. If the packet is absorbed, we then remove it from the simulation.

The advection of a packet by the fluid proceeds as follows. We define a momentum $p^\mu = A t^\mu + B u^\mu$, setting $A$ and $B$ by requiring that $p^\mu p_\mu =0$ and $p^\mu u_\mu = -\nu$. Then, we propagate the packet along the null geodesic defined by $p^\mu$ for a time $\Delta t'_{\rm adv} = \Delta t'_{\rm fl} (u^i p^t)/(u^t p^i)$. This propagates the packet over exactly the coordinate distance covered by the fluid during the time interval $\Delta t'_{\rm fl}$, but along a null geodesic instead of a timelike curve. Practically,
\beqn
\frac{A}{B} &=& -\frac{u_t + \sqrt{u_t^2 + g_{tt}}}{g_{tt}}, \\
B &=& \frac{\nu}{1-\frac{A}{B} u_t},\\
\frac{u^ip^t}{u^tp^i}&=& \frac{A+Bu^t}{Bu^t}.
\eeqn
As these equations are singular for a fluid with $u^i=0$, we only go through this procedure when $\delta_{ij} u^i u^j \geq 10^{-10}$. Otherwise, advection with the fluid motion simply keeps a packet at its current location. This algorithm does not exactly reproduce the diffusion of a packet in curved spacetime, but guarantees that packets experience the proper gravitational redshift, and are advected by the fluid. The choice to evolve along a null geodesic instead of along $u^\mu$ is done so that we can evolve the momentum of the packet according to Eq.~\ref{eq:EvP}, which assumes $p^\mu p_\mu=0$. We then propagate  the packet for $\Delta t_{\rm fl} - \Delta t_{\rm adv}$ along $t^\mu$. The vector $t^\mu$ is timelike, not null, but we make the approximation that the 4-momentum of neutrinos in the packet is constant during that step. In all of the above, the conversion between `primed' time intervals and `unprimed' time intervals is $\Delta t' = \Delta t/u^t$.

After this `advection' step, we move to the `free-streaming' step. We randomly draw a direction of propagation from an isotropic distribution in the fluid frame, as when emitting new packets, keeping the fluid frame energy $\nu$ constant. This generates a new fluid frame momentum for the packet, $p^{\mu}_{\rm free}$. There is now a subtlety in the determination of the time interval over which we let the packet free stream in the direction of $p^\mu_{\rm free}$. It is not simply $\Delta t_{\rm free} = f_{\rm free} \Delta t$,
because packets propagating in different directions should be propagated for the same time interval {\it in the fluid rest frame}. The correct time step for the free propagation of a packet is $\Delta t_{\rm free} = f_{\rm free} \Delta t [p^t / (\nu u^t)]$. With this weighting of the time step, the average propagation velocity of freely propagating packets is $u^i/u^t$, as desired. It is worth noting that because of this weighting, a given packet may be propagated to a time larger or smaller than initially requested. In our MC algorithm, there is however no particular requirement for the propagation of all packets to end at the same time in any given call to the MC algorithm, so this does not create any issue.

The last step in our approximate treatment of regions of high $\kappa_s$ is the most complex. Once a packet has been propagated, we need to determine what its momentum will be at the beginning of the next time step. We define the angles $(\theta_2,\phi_2)$ determining the orientation of that momentum vector in the fluid frame, in a spherical polar coordinate system whose polar axis is parallel to the momentum $p^{\mu'}_{\rm free}$. The momentum at the beginning of the next time step is then, in the fluid frame,
\beqn
p^{0'} &=& \nu  \nonumber \\
p^{1'} &=& \nu (-\sin{\theta_2}\cos{\phi_2}\cos\theta\cos\phi + \sin\theta_2 \sin\phi_2 \sin\phi  \nonumber \\
	&&+ \cos \theta_2 \sin \theta \cos \phi), \nonumber\\
p^{2'} &=& \nu (-\sin{\theta_2}\cos{\phi_2}\cos\theta\sin\phi - \sin\theta_2 \sin\phi_2 \cos\phi \nonumber \\
	&& + \cos \theta_2 \sin \theta \sin \phi), \nonumber\\
p^{3'} &=& \nu (\sin{\theta_2}\cos{\phi_2}\sin\theta + \cos\theta_2 \cos\theta),\nonumber
\eeqn
with $\theta$, $\phi$ the angles defining the orientation of $p^{\mu'}_{\rm free}$ in the fluid-frame orthonormal tetrad defined in Sec.~\ref{sec:def}.

By symmetry, we can always draw $\phi_2$ from a uniform distribution in $[0,2\pi]$. The simplest options for $\theta_2$ would be to either draw
$\cos\theta_2$ from a uniform distribution in $[-1,1]$, which effectively means that $p^{\mu'}$ is isotropically distributed in the fluid frame,
or to set $\theta_2=0$ and use the average direction of propagation of the packets away from the motion of the fluid as their final direction of propagation.
We will see that both options fail to reproduce solutions obtained without using any approximation in regions
of high $\kappa_s$. By inspection of three sets of  scattering experiments in which we scattered $\sim 10^8$ packets through optical depths 
$\tau_s = (3,10,30)$, however, we find that to high accuracy the distribution of $\theta_2$ only depends on the ratio $f_{\rm free}=\Delta t'_{\rm free}/\Delta t'$.
Furthermore, a good fit to the result of these experiments can be obtained by setting
\beq
\cos\theta_2 = B(f_{\rm free}) - [1+B(f_{\rm free})] \exp{\left[ r \ln{\left(\frac{B(f_{\rm free})-1}{B(f_{\rm free})+1}\right)}\right]},
\eeq
with $r$ drawn from a uniform distribution in $[0,1]$. To capture the fact that $p^{\mu'}$ should be isotropic for $f_{\rm free}=0$ and that $\theta_2=0$ for $f_{\rm free}=1$, we want $B(0)\rightarrow \infty$ and $B(1)\rightarrow 1$.
In practice, we fit $B$ to match our scattering experiments, and tabulate the result of those fits. We then interpolate $B$ linearly in $f_{\rm free}$. 
We use $B_i=B(0.05 i)$ with
\beqn
B_i = &(&1000,18.74,7.52,4.75,3.51,2.78, 2.32,2.00,\nonumber\\
	&&1.77,1.60,1.47,1.36,1.28,1.21,1.15,1.10,\nonumber\\
	&&1.07,1.04,1.019,1.0027,1.0000001)
\eeqn
and $i=0,1,...,20$. This approximation is tested in Sec.~\ref{sec:testKs}, and shows very good agreement with a full treatment of individual scatterings.
Nevertheless, it is only an approximation. When using this prescription, finite errors remain even in the limit of an infinite number of particles. Thus, despite the fact that 
the accuracy of this scheme is a small source of error compared to numerical errors at the grid resolution / number of MC packets that we can currently afford, we should keep in
mind that the coupled Moment-MC scheme only formally converges to an exact solution of Boltzmann's equation in the limit $\kappa_{\rm crit}\rightarrow \infty$ 
(i.e. when this approximate
method is used nowhere).

\subsection{MC-Moments Coupling}
\label{sec:MCM1Coupling}

The role of the MC evolution in our algorithm is to provide missing information about the distribution function of neutrinos to the evolution of the moments. In particular, we have shown in Sec.~\ref{sec:M1} that the moment scheme needs the Eddington tensor $\pi_{ij} = P_{ij}/E$, and the energy-averaged absorption and scattering opacities $\kappa_{a,s}$. In order to compute the evolution of the electron fraction of the fluid, we also need the number emission and absorption opacity, $\eta_N$ and $\kappa_N$. Indeed, the evolution equation for the composition of the fluid is
\beq
\partial_\mu (\rho Y_e \sqrt{-g} u^\mu) = -\sum_i {\rm sign}(\nu_i) \sqrt{-g} (\eta_N - \kappa_N J),
\eeq
where the sum is taken over all neutrino species, and ${\rm sign}(\nu_i)$ is $1$ for $\nu_e$, $-1$ for $\bar\nu_e$, and $0$ for heavy-lepton (anti)neutrinos.

We have already seen that the emissivity is simply $\eta = \sum_b \eta_b$, with $\eta_b$ the integrated emissivity within an energy bin. Similarly, $\eta_N = \sum_b 2\eta_b/(E_{b-1}+E_b)$. The other quantities will be derived from ratios of the MC moments $E_{MC}$, $F_{i,MC}$, $P_{ij,MC}$, $J_{MC}$, $(\kappa_a J)_{MC}$, $(\kappa_s J)_{MC}$, and $(\kappa_N J)_{MC}$. All of these moments are computed following the same strategy. Any MC moments $X_{\rm MC}$ is divided into three contributions
\beq
X_{\rm MC} = X_{\rm thick} + X_{\rm prop} + X_{\rm adv}.
\eeq
$X_{\rm thick}$ is the contribution from high opacity cells, as defined in Sec.~\ref{sec:largeKa}, i.e. cells where we do not evolve MC packets, and simply assume that neutrinos are in equilibrium with the fluid for the purpose of computing $(\kappa_a,\kappa_s,\kappa_N)$. $X_{\rm prop}$ is the contribution from packets propagating without any approximation, and from the portion of the time step $\Delta t_{\rm free}$ that a packet in a high-scattering region (i.e. treated using the approximation of Sec.~\ref{sec:highKs}) spends `free-streaming'. Finally, $X_{\rm adv}$ is the contribution from the portion of the time step $\Delta t_{\rm fl}$ that those same packets spend being explicitly advected with the fluid. 
This decomposition is necessary because MC packets are treated very differently (or entirely ignored) in regions of high opacity. It is not, however, equivalent to a more traditional separation between `trapped' and 'free-streaming' neutrinos. While neutrinos contributing to $X_{\rm thick}$ are trapped, neutrinos contributing to the other two components could be either trapped or free-streaming. In the limit of infinite resources, all packets should contribute to $X_{\rm prop}$.

Each of these three components relies on a different estimate of the stress-energy tensor, adapted to the treatment of neutrino packets in that region. 
For $X_{\rm thick}$, each energy bin which is not evolved by the MC algorithm is assumed to contribute
\beq
T^{\mu\nu}_{\rm thick} = \frac{\eta_b}{\kappa_{a,b} } \left(\frac{4u^\mu u^\nu+g^{\mu\nu}}{3}\right),
\eeq
as appropriate for neutrinos in equilibrium with the fluid. Here, $\kappa_{a,b}$ is the absorption opacity of the fluid
for neutrinos in the b$^{th}$ energy bin (computed at the center of the bin). For each propagating packet, we have
\beq
T^{\mu\nu}_{\rm prop} = N_k \frac{p_k^\mu p_k^\nu}{\sqrt{\gamma}\alpha p_k^t}  \delta^3(x^i-x^i_k),
\eeq
while for packets advected with the fluid, we use the approximation
\beq
T^{\mu\nu}_{\rm adv} \sim N_k \frac{\nu^2}{(\nu u^t)} \left(\frac{4u^\mu u^\nu+g^{\mu\nu}}{3\alpha \sqrt{\gamma}}\right)  \delta^3(x^i-x^i_k),
\eeq
which assumes an isotropic distribution of momenta in the fluid rest frame.
The full moment is then obtained by integrating over the proper volume $\sqrt{-g} \Delta V \Delta t$ of the cell (for the optically thick limit), or the propagating time $\Delta t$ of a packet. For example,
for the energy density $E_{MC}$,
\beqn
E_{MC} &=& \sum_{b,\rm thick} \sqrt{-g} \Delta V \Delta t  \frac{\eta_b}{\kappa_{a,b}} \left(\frac{4W^2-1}{3}\right) \\
&&+ \sum_{k,\rm prop} N_k (\alpha p_k^t)^2 \frac{\Delta t}{p^t_k}  \nonumber\\
&& + \sum_{k,\rm adv} N_k \nu \left(\frac{4W^2-1}{3}\right) \frac{\Delta t}{u^t}. \nonumber
\eeqn
The first sum is taken over the energy bins in which the cell is assumed to be optically thick, while the other two are taken over the paths of packets either propagating along null geodesics or advected with the fluid.
We note that there are several abuses of notation in the previous formula. A single packet undergoing scattering interactions may contribute to both the free-streaming and advected components of the stress-energy tensor during a single time step, with different $\Delta t$, and may contribute multiple time to the free-streaming component of the stress-energy tensor with different $\Delta t$ and $p^\mu$. Practically, the calculation is performed by letting each packet contribute to the computation of the moments whenever it is evolved in time. Packets evolved without approximations contribute once to $X_{\rm prop}$ for each time interval spent free-streaming in-between scattering or absorption events, while packets in high-scattering regions evolved using the approximation from Sec.~\ref{sec:highKs} contribute to both $X_{\rm prop}$ (for the period $\Delta t_{\rm free}$ spent approximately free-streaming) and $X_{\rm adv}$ (for the period $\Delta t_{\rm fl}$ spent being advected by the fluid). The optically thick regions are taken into account during the emission step, when we determine which cells are not evolved using the MC algorithm, and thus contribute to $X_{\rm thick}$.

The other moments can be obtained by taking different projections of the stress tensor, except for $(\kappa_N J)_{\rm MC}$ which is
\beqn
(\kappa_N J)_{\rm MC} &=&  \sum_{b,\rm thick} \sqrt{-g} \Delta V \Delta t  \frac{\eta_b}{\nu} \\
&&+  \sum_{k,\rm prop} N_k \kappa_a \nu \frac{\Delta t}{p^t_k} + \sum_{k,\rm adv} N_k \kappa_a \frac{\Delta t}{u^t}. \nonumber
\eeqn
With these moments at hand, we can compute $\pi_{ij} = P_{ij,\rm MC}/E_{\rm MC}$, and $\kappa_{a,s,N} = (\kappa_{a,s,N}J)_{\rm MC}/J_{\rm MC}$.
We also define the average number of packets per cell (multiplied by the proper time interval spent in that cell) $N_{\rm MC}$ as
\beqn
N_{\rm MC} &=&  \sum_{b,\rm thick} \sqrt{-g} \Delta V \Delta t  \frac{\eta_b}{\kappa_a E_p} \\
&& +\sum_{k,\rm prop} \frac{\nu\Delta t}{p^t_k} + \sum_{k,\rm adv} \frac{\Delta t}{u^t}. \nonumber
\eeqn

To time-average the MC moments over multiple time steps, and thus allow us to use a lower number of packets in the simulation, we continuously add to all of the above moments. Instead
of resetting the moments to zero at the beginning of a time step, we damp all moments (including $N_{\rm MC}$) using
\beq
X_{\rm MC} = X_{\rm MC}^0 \min{
\left[\exp{\left(-\frac{\Delta t}{u^t t_d}\right)},\frac{N_0 \Delta x_{\rm avg}}{N_{\rm MC}}\right]}.
\eeq
 Thus, $t_d$ is a minimum damping timescale in the fluid rest frame, while $N_0$ is the desired number of packets over which we average the neutrino distribution function,
 assuming that we can accumulate $N_0$ packets over a timescale shorter than $t_d$. The distance $\Delta x_{\rm avg} = (\sqrt{\gamma} \Delta x^1 \Delta x^2 \Delta x^3)^{1/3}$ provides a rough estimate of the propagation time across a cell. A large $N_0$ implies small statistical errors in the MC moments, but longer averaging timescales.
 Our standard choice in this manuscript is $N_0=100$, which should lead to $\sim 10\%$ relative errors in the MC moments. The actual accuracy of the method will however be better than this in optically thick regions, where more than 100 packets live in a cell at any given time. The impact of varying $N_0$ on the noise in the MC moments is explored in the tests of Sec.~\ref{sec:tests}. 

We also need a backup prescription for the computation of the moments when the MC algorithm does not have enough packets to provide any reliable information. Whenever we have a cell with $N_{\rm MC}<N_{\rm min} \Delta x_{\rm avg}$, all opacities are set to zero and the Eddington tensor is computed using the M1 closure, $\pi_{ij,M1}$. During the moment evolution, we also use $\pi_{ij,\rm M1}$ at cell faces if both neighboring cells have $N_{\rm MC}<N_{\rm min}\Delta x_{\rm avg}$. When that is the case, the characteristic speeds of the system are estimated assuming the nonlinear M1 closure. In this work, we use $N_{\rm min}=5$. We note that this prescription is necessary in many idealized problems in which neutrinos may be confined to a small portion of the grid, but probably less important in simulations of neutron star mergers, where neutrinos are present everywhere. 

The most natural interpretation of this `time-averaging' of the moments $X_{\rm MC}$ is that we damp the contribution of a packet to $X_{\rm MC}$ over the time scale 
needed for $N_0$ packets to cross a given grid cell, or over the user-prescribed time scale $t_d$ (whichever is smaller). The damping time scale thus varies from grid cell to grid cell, 
and over time. A disadvantage of this prescription is that we do not have absolute measurements of the moments, as $X_{\rm MC}$ cannot be easily normalized to provide a true 
average of $X$. In our algorithm, only ratios of the moments are meaningful quantities to use. Thus, we can use the Eddington tensor $\pi_{ij}=P_{ij}/E$, but not the energy
density $E$ itself. We have found, however, that a fixed damping timescale (which would allow us to properly normalize $X_{\rm MC}$) can lead to instabilities in optically thick
regions (if the time scale is too long), or to a highly inaccurate closure far away from the sources (if the time scale is too short). As any given cell can switch from the former to the 
latter over the course of a simulation, we need some adaptivity in the choice of the damping time scale. 

In SpEC, we evolve the MC packets before evolving the moments, and after evolving the fluid. When information from the MC evolution is required for the evolution of the moments, we always use the last-computed moments $X_{\rm MC}$. In regions
where more than $N_0$ packets cross a cell over a given time step, this means that $X_{\rm MC}$ is simply the average of $X$ over all packets evolved during the last MC time step.
In other regions, $X_{\rm MC}$ is an average of $X$ over the recent past of the system. This is a rather simple choice which, while appropriate for the type of conditions existing in
neutron star merger simulations, may be problematic in any system where the neutrino-fluid and/or moment-MC couplings are stiff enough that they have to be treated implicitly (e.g., possibly, in rapidly cooling optically thin regions existing in radiation-dominated accretion disks).

Finally, we will show in the code tests that, at finite resolution, the moment and MC algorithms can become inconsistent. Whether this will be a significant issue for any given application remains an open question, as is the best method to avoid this issue. 
To illustrate a potential way around this issue, we implement a fairly simple-minded method which attempts to switch to a pure moment evolution when the moment and MC schemes
become so inconsistent that using the MC moments may be meaningless. We define a deviation between the MC and moment algorithms as 
\beqn
(\delta F)^2 &=& \gamma^{ij} \left(\frac{F_{i,\rm MC}}{E_{\rm MC}}-\frac{F_{i}}{E}\right) \left(\frac{F_{j,\rm MC}}{E_{\rm MC}}-\frac{F_{j}}{E}\right),\\
\epsilon &=& 0.5\left[1+\rm{atan}\left(\frac{\min{(\delta F,1)}-0.5}{0.1}\right)\right].
\eeqn
The Eddington tensor is then defined as
\beq
\pi_{ij} = \epsilon \pi_{ij,\rm M1} + (1-\epsilon) \pi_{ij,\rm MC}.
\eeq
This introduces a lag in the reaction of the evolution equations to a change in the closure, but also provides additional stability.
The approximate characteristic speeds are similarly corrected using $c_{\pm}=  \epsilon c_{\pm}^{MC} \pm (1-\epsilon) c^{M1}$ with $c^{M1}$ the largest characteristic speed (in absolute value) for the evolution of the moments, assuming the M1 closure. Potential alternatives to this prescription includes promoting $\delta F_i = \left(\frac{F_{i,MC}}{E_{MC}}-\frac{F_{i}}{E}\right)$ to an evolved variable, and using constraint damping to drive that variable to zero, or using high-order/low-order methods in which $\delta F_i$ is used to compute a stabilizing source term in the evolution equations for $F_i$. We do not investigate these other methods in this work. We note that the method implemented here does
not stop convergence towards a true solution of Boltzmann's equations with increased numerical resources, but it may make these solutions numerically unstable. At the very least, 
we note in our existing tests that the convergence properties of the uncorrected coupled Moment-MC scheme are better than those of the corrected scheme in all but one configurations, in which both schemes encounter significant problems (the relativistically boosted emitting sphere of Sec.~\ref{sec:movingsph}).

\subsection{Fully coupled general relativistic radiation hydrodynamics}
\label{sec:fullCoupling}

Although in this work we never evolve Einstein's equations nor the general relativistic equations of hydrodynamics, our MC-Moments algorithm is already fully integrated into SpEC. The coupled general relativistic radiation hydrodynamics system is put together as follow. Einstein's equations and the general relativistic equations of hydrodynamics are evolved as usual in SpEC, using a third-order Runge-Kutta time stepping method. Einstein's equations are evolved on a pseudospectral grid, and the general relativistic equations of hydrodynamics on a finite volume grid (see~\citealt{Duez:2008rb,Foucart:2013a}). The source terms are communicated between these grids at the end of each full step of the Runge-Kutta algorithm. Values at intermediate steps are obtained by first-order extrapolation in time. The neutrinos are treated using operator splitting. At the end of a metric/fluid time step $\Delta t$, we evolve the two-moments equations by $\Delta t$. During the neutrino timestep, we also modify the fluid variables to take into account the impact of neutrino-matter interactions on the energy density, momentum, and composition of the fluid. 

The MC algorithm can, in theory, be called at arbitrary time intervals. In particular, if the averaging timescale is long compared to $\Delta t$, there is a priori no need to call the MC evolution at each timestep. However, the main cost of the MC evolution is the propagation of neutrino packets, and neutrino packets propagate at the speed of light between neighboring cells.
We argued in Sec.~\ref{sec:propagation} that this leads to a limit on the time step comparable to the Courant condition limiting the time step for the evolution of the fluid and moments of $f_{(\nu)}$.
Accordingly, there is very little to gain by using a different timestep for the fluid evolution and for the MC algorithm. One exception is if the pseudospectral grid used to evolve the metric is much finer than the finite volume grid used for the fluid/neutrinos, and sets the timestep for the fluid/metric evolution. In that case, it may be beneficial to only call the MC algorithm every $n$ steps of the metric/fluid evolution, with $n$ chosen so that $n \Delta t \lesssim \Delta x/3$. As the code tests below do not evolve Einstein's equations, we do not encounter such a case here, and set $n=1$.

Each step of the MC algorithm proceeds, in our current implementation, as follow: 
\begin{itemize}
\item Packets are created as described in Secs.~\ref{sec:emission} and~\ref{sec:largeKa}, as needed to represent neutrino emission and the sampling of an equilibrium distribution of neutrinos in optically thick cells at the boundary of the MC domain. Each packet is initially owned by the processor which owns the fluid cell in which the packet is created. In optically thick cells which are not evolved with the MC algorithm, we add to the MC moments the contribution of neutrinos in thermal equilibrium with the fluid.
\item Packets are propagated, scattered, and absorbed following the methods of Secs.~\ref{sec:propagation} or~\ref{sec:highKs}. The contribution to the MC moments of each packet is computed at this time.
\item The MC moments are communicated to the ghost zones of the fluid grid, so that the evolution of the moments can use values of the MC moments in ghost zones.\footnote{Ghost zones are buffer grid cells neighboring the cells evolved by a given processor. They are not evolved, but they are needed for high-order interpolation from cell centers to cell faces. For the fifth-order WENO reconstruction used here, we need 3 buffer cells along each dimension. The values of the fields in those cells have to be overwritten by their values on the processor in which that region of the grid is actually evolved.}
\item All packets which ended their evolution in a ghost zone of the fluid grid are transferred to the processor which evolves the region of the grid corresponding to that ghost zone.
\end{itemize}
The parallelization method chosen here, in which packets are simply owned by the processor that also owns the fluid cell where the packets reside, is highly suboptimal. Indeed, the distribution of packets across grid cells is very uneven, with most packets residing in the most optically thick cells evolved by the MC algorithm. We expect that, for efficient 3D evolution of neutron star mergers, a more advanced parallelization scheme will be necessary. In this work, however, we focus solely on the stability and accuracy of the mixed MC-Moments algorithm, and do not attempt to implement a more advanced parallelization scheme.

We also note that the methods described here are mostly useful for systems in which we can rely on time-averaged moments to reduce the cost of the simulation. This makes their application to neutron star merger simulations very promising, but their use in radiatively dominated accretion disks more uncertain. In accretion disks, we may have optically thin regions in which the cooling time scale is shorter than the evolution time step, and radiatively dominated regions in which small errors in the neutrino pressure can create large errors in the fluid evolution. The savings provided by time-averaging the moments may be negligible in those regimes. A potential alternative would be to incorporate in our scheme implicit MC techniques, as developed by~\cite{Roth2015} -- although how this would interface with the rest of our algorithm is highly speculative at this point.

\section{Code Tests}
\label{sec:tests}

\subsection{Packet propagation}

\begin{figure}
\includegraphics*[width=0.49\textwidth]{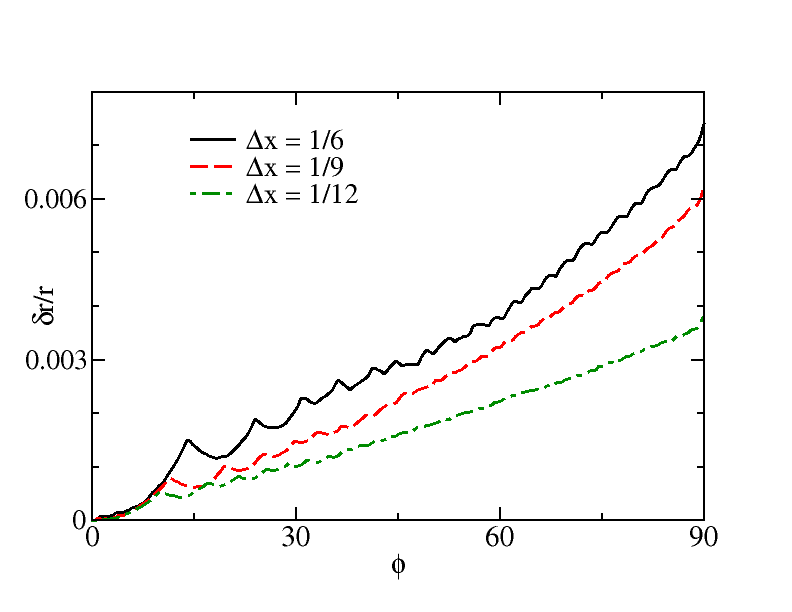}
\caption{Numerical error in the propagation of a single packet around a black hole of unit mass. The packet is initialized at $x^i=(0,4,0)$, with
a momentum such that the packet initially propagates along the +x direction. While the low-order methods used to propagate packets cause slow
convergence of the solution, the amplitude of the numerical error is quite small.}
\label{fig:TrajectoryError}
\end{figure}

As a first code test, we consider the propagation of a single neutrino packet in curved spacetime. This allows us to verify that free-streaming packets follow
null geodesics, and to obtain an estimate of the numerical error due to the use of low-order methods for packet propagation along geodesics. We consider a non-spinning black hole of mass $M_{\rm BH}=1$, and use the Kerr-Schild metric $g_{\mu\nu} = \eta_{\mu\nu} + 2 l_\mu l_\nu/r$, with $l_\mu = (1,x_i/r)$. The packet is initialized at $x^i=(0,4,0)$,
with a fluid rest frame energy $\nu=1$ and a 4-momentum set by the additional requirements that $p^2=p^3=0$. We follow the packet up to the point at which $x^2=0$. We show the error in the location of the packet as a function of the azimuthal angle $\phi=atan{(x^1/x^2)}$, for grid spacings $\Delta x = (1/6,1/9,1/12)$ and time steps $\Delta t = (0.075,0.05,0.0375)$. This is a relatively low resolution compared to what is typically used in simulations of NSNS and NSBH mergers. Yet, we see on Fig.~\ref{fig:TrajectoryError} that even for a packet initialized fairly close to the black hole, the numerical error remains small, $\lesssim 0.5\%$.
Similar errors are observed for the conserved energy and angular momentum of the packets, $p_t$ and $p_\phi$. We thus conclude that the low-order propagation methods used in this work are unlikely to be a significant source of error in merger simulations. We also note that the systematic nature of the error (i.e. the fact that the error continuously grows over time) is due in part to the monotonicity of all derivatives of the curve representing the packet's orbit, and in part to the divergence of the trajectory of massless particles moving on close, outspiraling orbits around a black hole, once a small numerical error has been introduced.

The convergence of the numerical solution to the expected analytical trajectory is quite slow. In fact, formal convergence is only observed at very early times, for $\phi \lesssim 5^\circ$. This indicates that to obtain reliable error estimates for the propagation of neutrino packets, higher order methods would be required. However, as the actual error in the evolution of the packets observed here is negligible compared to other sources of error in current general relativistic radiation hydrodynamics simulations, we do not consider it worthwhile to go beyond the cheap, low-order methods implemented in our current code.

\subsection{Single Beam in Curved Spacetime}

\begin{figure}
\includegraphics*[width=0.49\textwidth]{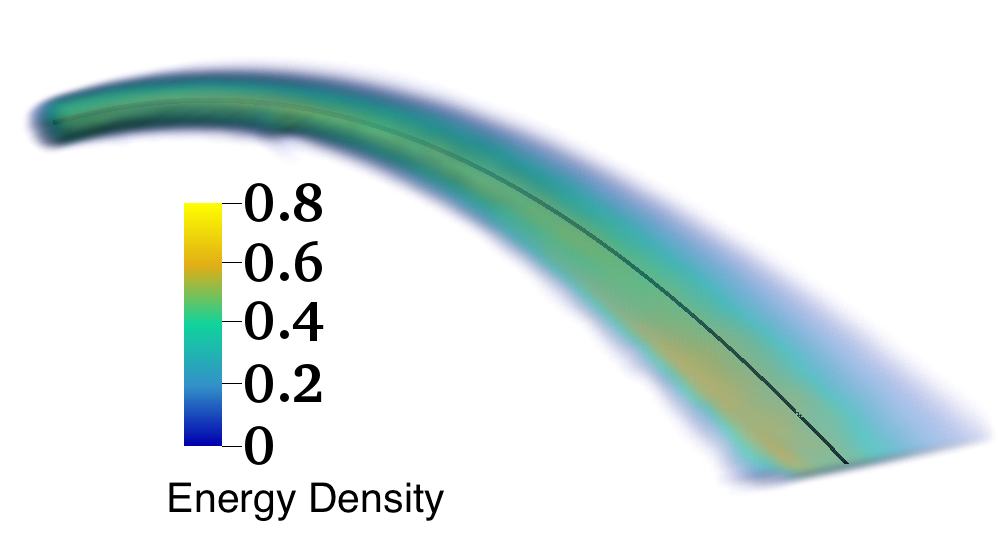}
\caption{Beam distorted by the gravitational potential of a non-spinning black hole of unit mass. All neutrinos are emitted from a small sphere of radius $r_s=0.2$ around $x_c=(0,4,0)$,
with initial motion parallel to the X-axis (top left part of the plot). With that choice, all neutrinos pass through the $(x^2=0, x^3=0)$ line (bottom right part of the plot). At that point, the height of the beam goes down to a single grid cell. The solid black line is the analytical result for the null geodesic at the center of the beam. The color scale shows the neutrino energy density $E$. We show the beam at a time $\Delta t=18 GM_{\rm BH}/c^3$ after the beginning of the simulation. The result is, however, nearly identical for all times after the beam crosses the outer boundary of our grid.}
\label{fig:BHbeam}
\end{figure}

We now move to the coupled MC-Moments scheme, with a similar setup. Instead of emitting a single packet at $x^i=(0,4,0)$, we create packets in a small sphere of radius $r_s=0.2$ around that point. All packets are still initialized with a momentum such that $p^2=p^3=0$, $\nu = 1$.
The emissivity is set to $10^6$ packets per unit volume (with $G=c=1$).  With the grid spacing $\Delta x = 0.057$ and time step $\Delta t = 0.025$ used in our simulation, this corresponds to only $\sim 4.6$ packets per time step in each cell within the emitting region. As opposed to the previous test, we evolve the lowest two moments of the neutrino distribution function, with a closure provided by the MC scheme. This setup allows us to test two important aspects of the code: the evolution of a single beam in curved spacetime, and the collapse of that beam into the $x^3=0$ plane as it crosses the $x^2=0$ plane (as all packets emitted with $p^2=p^3=0$ on the $x^1=x^3=0$ axis pass through the $x^2=x^3=0$ axis).

A volume rendering of the beam is shown on Fig.~\ref{fig:BHbeam}. We see that the beam follows the expected null geodesic, and that the height of the beam becomes of the order of $\Delta x$ on the $x^2=x^3=0$ line. This does not happen when using the M1 closure for the evolution of the moments: with that closure, a convergent beam is partially supported by unphysical radiation pressure, and the evolution of the beam converges to an erroneous solution. With the MC closure, we can collapse the beam to a height of a single grid cell!
In Fig.~\ref{fig:BHbeam}, the $x^2=0$ plane is also the outer boundary of our domain. We find similar results when placing that boundary at $x^2=-4$.

\subsection{Crossing Beams}

\begin{figure}
\includegraphics*[width=0.15\textwidth]{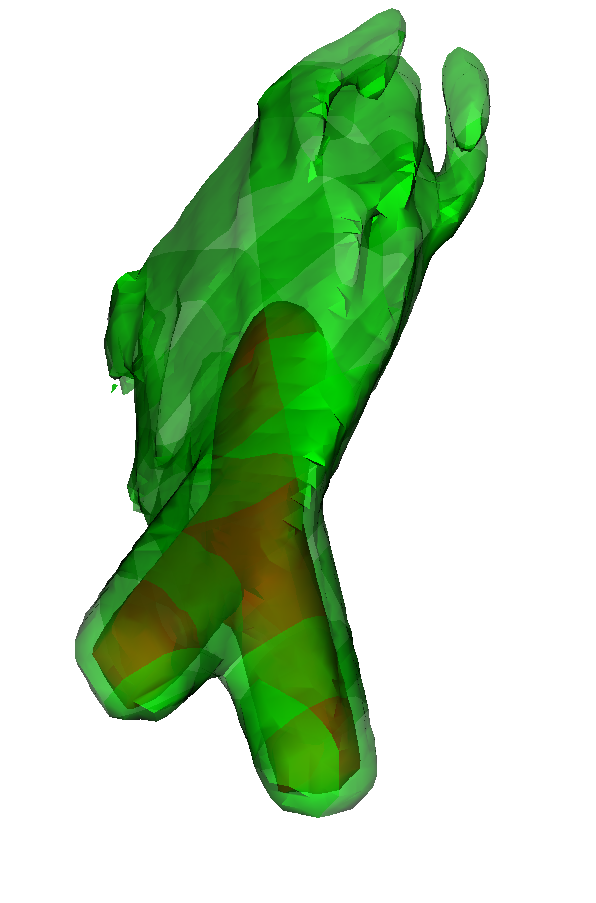}
\includegraphics*[width=0.15\textwidth]{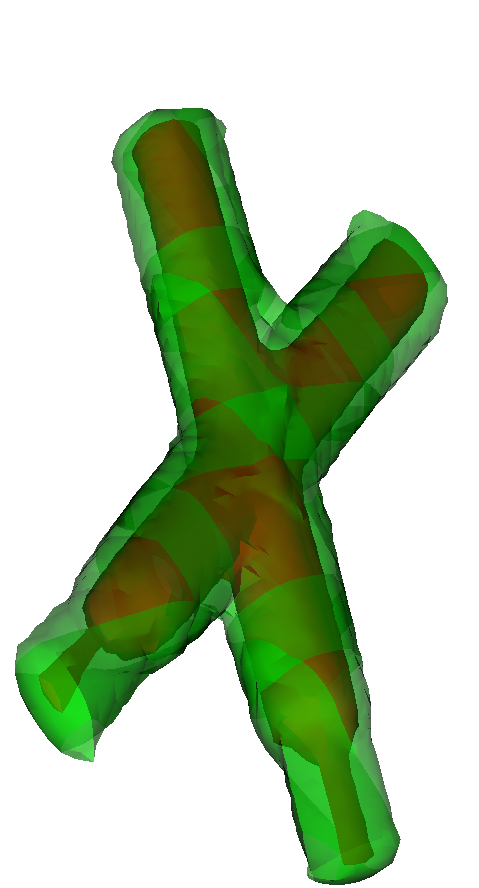}
\includegraphics*[width=0.15\textwidth]{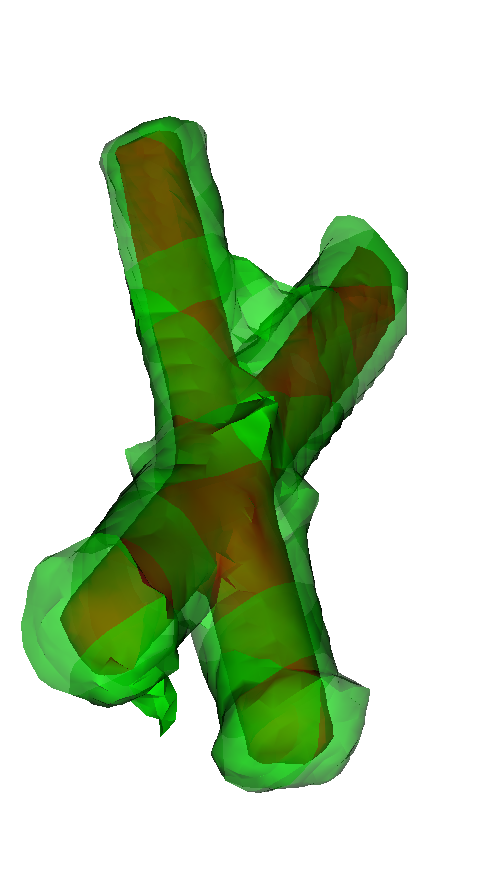}
\caption{Two beams of neutrinos emitted from spheres of radii $r_s=0.3$ and centers $c_1=(-2,-1,0)$, $c_2=(-1,-2,0)$, towards the origin. We show contours at 10\% and 1\% of the maximum energy density $E$. Three different versions of the test are shown, with the beams shot upwards from the bottom part of the plot: {\it Left}: M1 closure, showing the unphysical collision of the two beams, {\it Center}: MC closure, {\it Right}: MC closure, but with a correction from the M1 closure when the M1 and MC moments disagree (see Sec.~\ref{sec:MCM1Coupling}). We visualize our `medium' resolution, at $t=10GM/c^3$.}
\label{fig:CrossingBeams}
\end{figure}

For our next test, we consider a well-known problem where the standard M1 closure fails spectacularly. We set up two coplanar beams of neutrinos, emitted from two spheres of radii $r_s=0.3$ and centers $c_1=(-2,-1,0)$, $c_2=(-1,-2,0)$. The beams are emitted towards the origin, and cross each other there. With the M1 closure, the beams collide and then propagate along the direction of their average momentum (see e.g.~\citealt{FoucartM1:2015}, or Fig.~\ref{fig:CrossingBeams}) -- an obvious manifestation of the problems encountered by the M1 closure for converging beams. 

With the MC closure, on the other hand, the beams properly cross, as shown in Fig.~\ref{fig:CrossingBeams}. Due to the use of time-averaged moments, and the finite time required for the moment evolution to react to the passage of packets in the MC evolution, our evolution is not entirely free of artifacts: numerical diffusion at the level of a few percents of the maximum density of the beams is clearly visible in Fig.~\ref{fig:CrossingBeams}, and part of the beam energy is reflected by the crossing region. Yet, in this test, the time-averaged MC closure provides  results which are far superior to the M1 closure.

\begin{figure*}
\includegraphics*[width=0.9\textwidth]{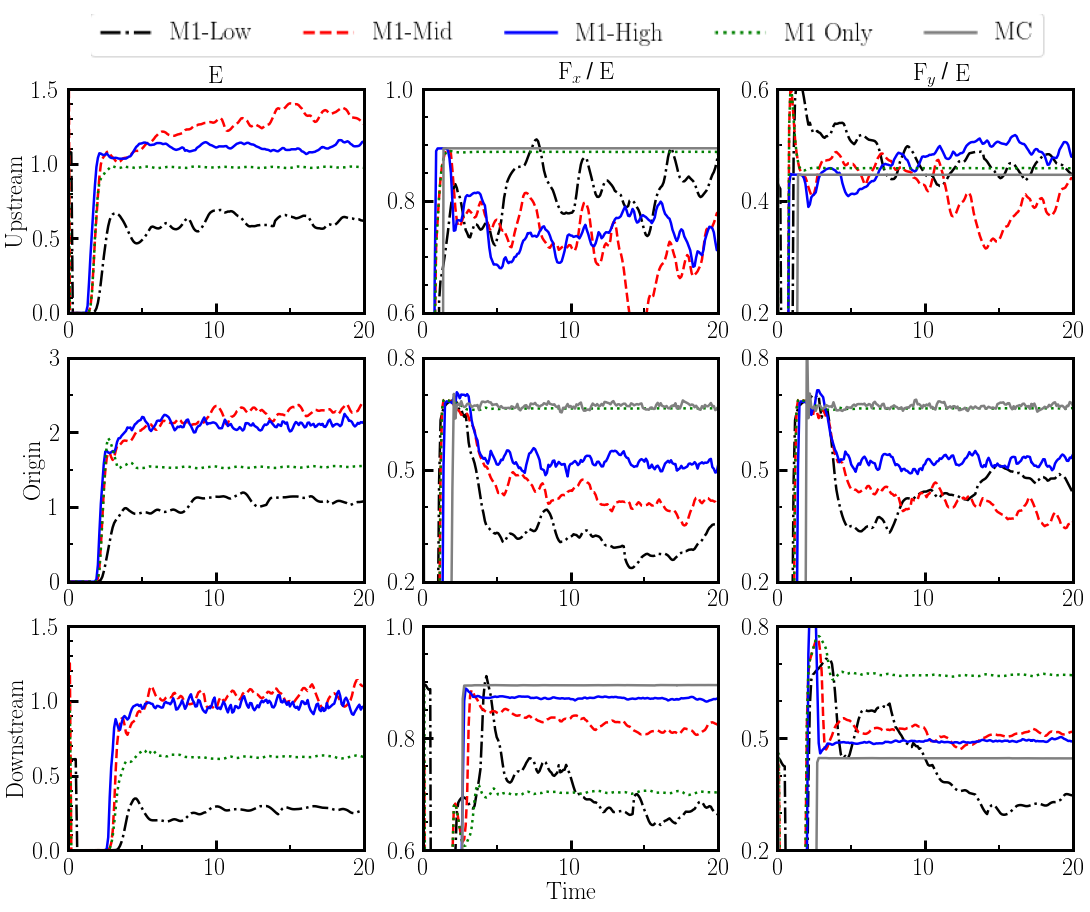}
\caption{Energy density $E$ ({\it left}), and flux density along the x-axis $F_x$ ({\it middle}) and y-axis $F_y$ ({\it right}) in the crossing beam test. $E$ is normalized to the correct energy density at the center of a single beam, while $F_i$ is normalized by $E$. We show results upstream of the crossing region ({\it top}), at the center of the crossing region ({\it center}), and downstream of the crossing region ({\it bottom}). In each panel, we plot the moments as evolved in the two-moment scheme at three resolutions (from low to high, dot-dashed black curve, dashed red curve, and blue curve) with the closure provided by the MC evolution. We also plot the same moment when the closure is provided by the analytical M1 prescription (dotted green curve), at the middle resolution. Finally, we show the moments measured in the MC evolution (grey curve, except for $E$, as that information is not available in our code). We see that the pure M1 scheme avoids reflections at the crossing point, thus obtaining the correct answer upstream of that point, but is generally inaccurate downstream of the crossing point. With the MC closure, the solution converges to the correct answer everywhere.}
\label{fig:CrossingBeamEF}
\end{figure*}

\begin{figure*}
\includegraphics*[width=0.9\textwidth]{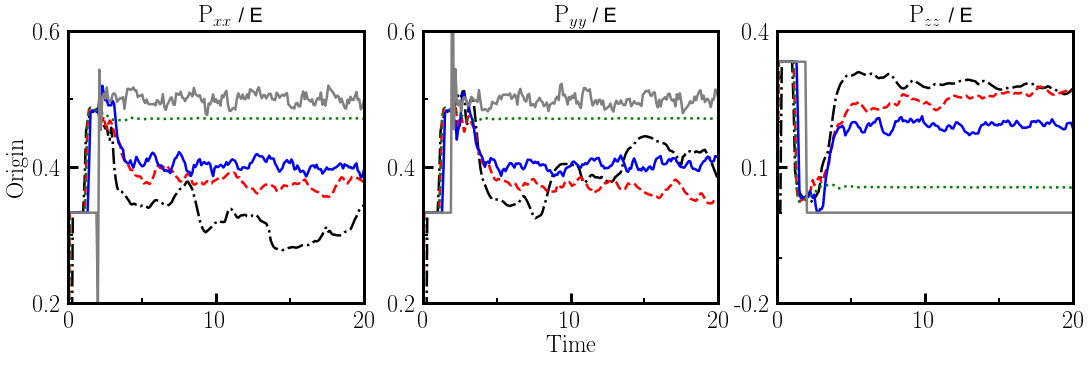}
\caption{Same as Fig~\ref{fig:CrossingBeamEF}, but for the diagonal components of the Eddington tensor $P_{xx}/E,P_{yy}/E,P_{zz}/E$. Note that here the fully coupled code uses the Eddington tensor as measured in the MC code (grey curves), while the pure M1 code uses the M1 closure (green curve). The other curves are provided for illustrative purposes only. We see that the M1 closure assumes a non-zero $P_{zz}$ in the crossing region, while the MC code properly measures $P_{zz}=0$.}
\label{fig:CrossingBeamP}
\end{figure*}

To better understand what happens in this test, let us look at moments of the neutrino distribution function at the origin (where the beams cross) as well as at points $(-0.6,-0.3,0)$
and $(0.6,0.3,0)$ located, respectively, upstream and downstream of the crossing point, at the center of one of the beams. We measure the energy density $E_{\rm M1}$ and normalized flux density $f_{i,\rm M1}=F_i/E$ as evolved in the moment scheme, the Eddington tensor $\pi_{ij,{\rm M1}}=P_{ij}/E$ obtained by applying the M1 closure on $E,F_i$, and the ratio of moments $f_{i,\rm MC}=(F_i/E)_{\rm MC}$ and $\pi_{ij,\rm MC}=(P_{ij}/E)_{\rm MC}$ computed from the MC packets. In our fully coupled algorithm, $\pi_{ij,{\rm M1}}$ is only used in regions where no MC packets are present, and $f_{i,\rm MC}$ is not used at all. The first is useful to determine when the M1 closure fails, while for the simple problem considered here, $f_{i,\rm MC}$ and $\pi_{ij,\rm MC}$ are both exact solution to the radiation transport problem, up to sampling noise in the region where the two beams interact (in this simple test, MC packets propagate in a straight line, are never absorbed or scattered, and have the same momentum if they belong to the same beam).

Fig.~\ref{fig:CrossingBeamEF} shows $E$ and $f_i$ at these three points. We use three resolutions with grid spacing $\Delta x =(0.25,0.12,0.06)$, and set the number of MC packets that we average moments over to $N_0=50,100,200$. The energy of each packet ($E_p$) is decreased by a factor of $16$ at each resolution. At the point of highest neutrino density (the origin), we have $\sim(10, 20,40)$ packets per grid cell and thus damping time scales for the averaging of the moments $ \sim (1.25,0.6,0.3)$. The longest damping time scale that we allow is $t_d=(10,5,2.5)$. That time scale is thus in use in regions where the beam intensity is $\lesssim 25\%$ of the intensity at the center of the beam. We also show results using the analytical M1 closure. We see that upstream of the crossing point, even though the energy density in the MC-Moments scheme converges to the correct solution, the pure M1 scheme performs better than the coupled scheme. This is because numerical errors in the pure M1 scheme do not lead to any reflection in the crossing region. Downstream of the shock, on the other hand, the pure M1 scheme does not converge to the physical solution, while the coupled scheme provides good results as soon as we have $\sim 5-10$ grid points across the beam. By varying independently $E_p$ and $N_0$, we also find that the damping timescale and the number of packets used to compute the moments have minimal effects on the quality of the results, beyond the expected changes in the sampling noise in the MC moments. The error made by the M1 closure in the crossing region is illustrated on Fig.~\ref{fig:CrossingBeamP}, which shows the Eddington tensor provided by the MC and M1 closures. In that region, the M1 closure assumes a combination of free-streaming neutrinos along the average direction of the two-beams, and of an isotropic distribution of neutrinos. One consequence of this assumption is that $P_{zz}\neq 0$ in the M1 closure, while it vanishes exactly in the MC closure (there is no sampling noise because all neutrino packets have $p^3=p^z=0$ at all times).

We can also comment on the impact of replacing the MC closure with a combination of the MC and M1 closure, when the M1 and MC moments disagree (see Sec.~\ref{sec:MCM1Coupling}). Fig.~\ref{fig:CrossingBeams} shows that this mixed approach has slightly larger errors in the crossing region that the pure MC closure.
It avoids reflection of the beams through the source, but without improving the quality of the solution upstream of the crossing region. Most importantly, however, the steady-state
solution obtained with the mixed closure does not converge to the true steady-state solution, i.e. the solution does not noticeably improve from what is shown on Fig.~\ref{fig:CrossingBeams} at high resolution. The code converges to the correct solution at very early times, but that solution does not appear to be stable for the resolution
and number of packets used in our tests. The mixed closure performs better than the M1 closure, but clearly worse that a direct use of the MC closure in this test.
Finally, we note that the performance of the M1 and mixed closure becomes worse as the angle between the two beams increase, while the MC closure provides good results for all beam orientations that we have attempted -- including for two overlapping beams propagating in opposite directions.

\subsection{Diffusion through a region of high scattering opacity}
\label{sec:testKs}

\begin{figure}
\includegraphics*[width=0.49\textwidth]{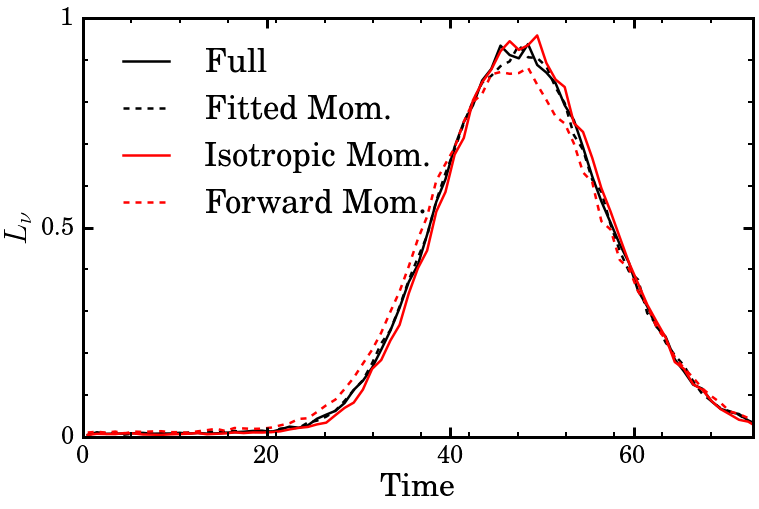}
\caption{Luminosity of the MC packets leaving the computational domain, as a function of time, for a Gaussian pulse advected in a moving medium.
The medium has $\kappa_a=0$ and $\kappa_s=100$, with $\Delta t=0.05$, so that the code can use the simplified evolution of neutrino packets in high-$\kappa_s$ regions.
The different curves represent results without using the approximate method (Full), with our preferred approximate method in which the final momenta are fitted to the results
of scattering experiments (Fitted Mom.), and with momenta chosen either isotropically in the fluid frame (Isotropic Mom.) or always in the direction of propagation of the packet away from the fluid motion (Forward Mom.). The latter is the only method which doesn't quite reproduce the correct luminosity.} 
\label{fig:scatLum}
\end{figure}

\begin{figure}
\includegraphics*[width=0.49\textwidth]{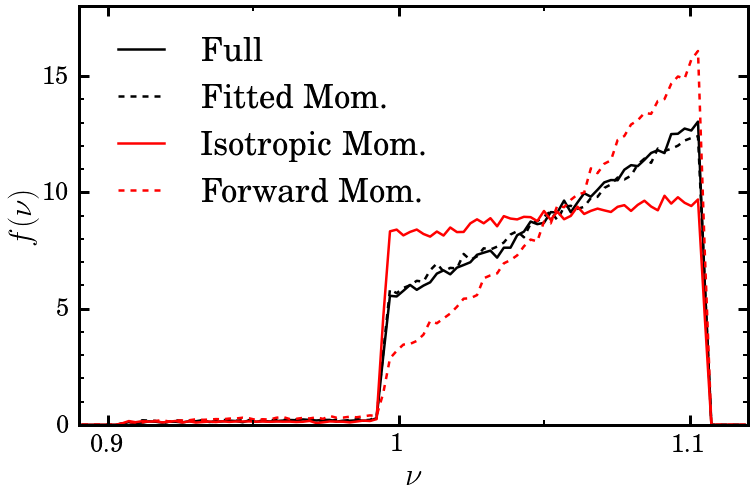}
\caption{Spectra of the MC packets for the same test as in Fig.~\ref{fig:scatLum}. The energy is normalized to the energy at which packets are emitted in the fluid rest frame. The vertical axis $f(\nu)$ is the probability distribution of neutrino energies. We see that the various approximate methods to choose the final momenta of neutrino packets in our approximate treatment of scattering events lead to significant differences in the spectra, with only the method in which the final momenta are fitted to the results of scattering experiments providing a good match to the correct solution.}
\label{fig:scatSpec}
\end{figure}

One of the most complex part of our coupled MC-Moments algorithm is the approximate treatment of regions of high scattering opacities, discussed in Sec.~\ref{sec:highKs}.
A standard test for radiation transport in optically thick regions is the advection of a Gaussian pulse in a fluid with constant velocity $u^\mu=W(1,v,0,0)$, on a Minkowski background and with opacities $\kappa_a=0$, $\kappa_s \Delta x \gg 1$. In our case, this is however a fairly trivial test: the evolution of the moments is already corrected to reproduce the diffusion limit when $(\kappa_s+\kappa_a) \Delta x \gg 1$, and does not use the MC closure in that regime. While our code does maintain a steady pulse profile comoving with the fluid, as expected, this is not a particularly useful test of our algorithm. We simply note that the main source of error in this test comes from the reconstruction of the moments on cell faces. That error is negligible when using the high-order WENO5 reconstruction in the evolution of the moments, but can be significant for more dissipative reconstruction methods.

A more interesting test of our algorithm is when $\kappa_s \Delta t \gtrsim 3$: in that case, the evolution of the moments uses a weighted average of the MC closure and of the diffusion approximation and, more interestingly, the MC algorithm uses the complex method approximating the propagation of neutrino packets through a region of large $\kappa_s$ described in Sec.~\ref{sec:highKs}. 

To test this regime, we consider a simple cartesian grid with spacing $\Delta x=0.2$, and initialize a Gaussian pulse of width $\Delta = 0.5$ at the origin. The background fluid has a velocity $u^\mu=W(1,0.1,0,0)$ and an energy-independent $\kappa_s = 100$ (except within 3 cells of the boundary, as well as for $x>4.5$, where $\kappa_s=0$). The computational domain is defined by $-2.6<x<7$, $-2.6<y<2.6$, $-2.6<z<2.6$. We use a time step $\Delta t = 0.05$. We find that the advection and diffusion of the pulse is well captured, fairly independently of the details of the algorithm used for regions of high $\kappa_s$. This is most likely because the evolution of the moments still relies largely on the diffusion limit for $\kappa_s \Delta x = 20 \gg 1$. 
On the other hand, we find that using a well-calibrated prescription for the choice of the momentum of the neutrino packets at the end of their approximate propagation through high $\kappa_s$ regions plays an important role in properly capturing the spectrum of the neutrinos escaping that region. 

We first focus on the properties of the MC packets escaping the region of high $\kappa_s$. We measure the number of packets leaving the computational grid which, when combined with the known energy of a packet, provides us with a luminosity $L_{\nu}$ {\it according to the MC algorithm}. We also measure the energy spectrum of these escaping MC packets. These measurements are possible because our algorithm writes on disk the properties of all packets leaving the computational grid, or of a subsample of these packets, to allow 
post-processing of that information. We consider the evolution of the moments later in this section. The spectral information is particularly important in the context of neutron star 
mergers: the average energy of the neutrinos escaping high-density regions and the shape of the neutrino spectrum have a significant impact on the absorption rate of neutrinos in 
low-density regions, and on the evolution of the composition of the fluid. To test the approximate treatment of high-scattering regions described in Sec.~\ref{sec:highKs}, we consider
4 different algorithmns:
\begin{enumerate}
\item {\bf Full} scattering treatment: every scattering event is treated individually, causing the code to re-draw the momentum of the neutrinos from an isotropic distribution 
in the fluid rest frame. This method is expensive in high-$\kappa_s$ regions, and used here only to test the validity of the other algorithms.
\item {\bf Fitted momenta} scattering treatment: packets are evolved through high-scattering regions using the prescriptions of Sec.~\ref{sec:highKs}. The final
momentum of the packets and the distance that the packets move away from the average motion of the fluid are both drawn from distributions fitted to scattering experiments. This is our preferred method in high-$\kappa_s$ regions.
\item {\bf Isotropic momenta} scattering treatment: the final momentum of the packets is drawn from an isotropic distribution in the fluid rest frame, instead of from a fit to scattering experiments. The motion of the packets is treated as in the previous case. This is a good approximation for packets which do not move much with respect to the fluid over a given time step.
\item {\bf Forward momenta} scattering treatment: the final momentum of the packets is in the same direction as the motion of the packet in the fluid rest frame. This is a good approximation for packets that largely avoid scattering events during a given time step (i.e. packets with average velocity with respect to the fluid close to the speed of light).
\end{enumerate}
The last two methods provide us with an estimate of the cost of abandoning the most complex part of the algorithm from Sec.~\ref{sec:highKs}, the final choice of the momentum of
the packets, in favor of simpler (but more approximate) methods.

Fig.~\ref{fig:scatLum} shows the flux of MC packets across the domain boundary as a function of time for these 4  algorithms. We see that only the {\it Forward} method shows a (small) error in the flux of neutrino packets. All methods also agree well with the measured flux in the evolution of the moments. 
Fig.~\ref{fig:scatSpec} shows the energy spectrum of the packets  leaving the grid. Now, only the {\it Fitted} method properly matches the evolution in which we do not use any approximation. The {\it Forward} method overestimates the impact of the Doppler shift due to the motion of the fluid, while the {\it Isotropic} method underestimates that effect. This 
can be understood from the fact that most packets are advected with the fluid and leave the region of high-$\kappa_s$ on the $x>0$ side of that region. These escaping packets 
experience relativistic beaming, and are thus preferentially moving in the direction maximizing the Doppler shift. The {\it Isotropic} prescription does not capture relativistic beaming, 
while the {\it Forward} prescription overstates its effect.
Accordingly, we consider it worthwhile to use the more complex (but not significantly more expensive) method derived in Sec.~\ref{sec:highKs} to evolve neutrino packets in high-$\kappa_s$ regions.

\begin{figure}
\includegraphics*[width=0.49\textwidth]{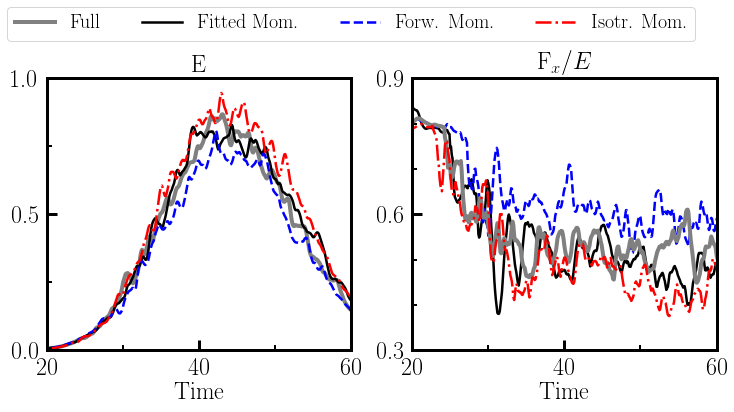}
\caption{Normalized energy density $E$ and normalized flux $F_x/E$ in the same test as in Figs.~\ref{fig:scatLum}-\ref{fig:scatSpec} and for the same 4 
algorithms. The moments are measured at point $(4.6,0,0)$, right outside of the high-scattering region and in the direction of motion of the background fluid. 
The moments confirm the picture of Fig.~\ref{fig:scatSpec}: different algorithms lead to minor changes in energy density, and the {\it Forward} (resp. {\it Isotropic}) algorithm overestimates (resp. underestimates) the effect of relativistic beaming.}
\label{fig:scatMom}
\end{figure}

In this test, local values of the moment do not provide quite as much information as a study of the MC packets. In Fig.~\ref{fig:scatMom}, we show the energy and flux density
of neutrinos in the evolution of the moments, right outside of the high-$\kappa_s$ region in the direction of motion of the background fluid. The moments are consistent with the
information gleaned from the study of the MC packets. The energy density is only mildly affected by the choice of algorithm, while relativistic beaming (which affects $F_x/E$) is only properly captured by the {\it Fitted} algorithm.

\begin{figure}
\includegraphics*[width=0.49\textwidth]{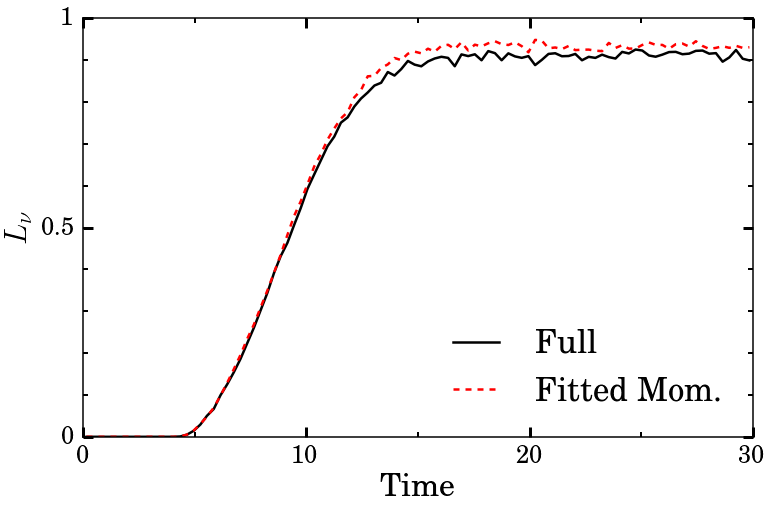}
\caption{Same as Fig.~\ref{fig:scatLum}, but for a spherical shell $2.5<r<5$ with $\eta =1$, $\kappa_a=1$, and $\kappa_s=100$ around a non-spinning BH. We only show our preferred method for the determination of the momentum of packets, and the result of an evolution without using any approximation in high-scattering regions. The luminosity is normalized for ease of visualization. The approximate method overestimates the luminosity by $\sim 2\%$.} 
\label{fig:scatBHLum}
\end{figure}

\begin{figure}
\includegraphics*[width=0.49\textwidth]{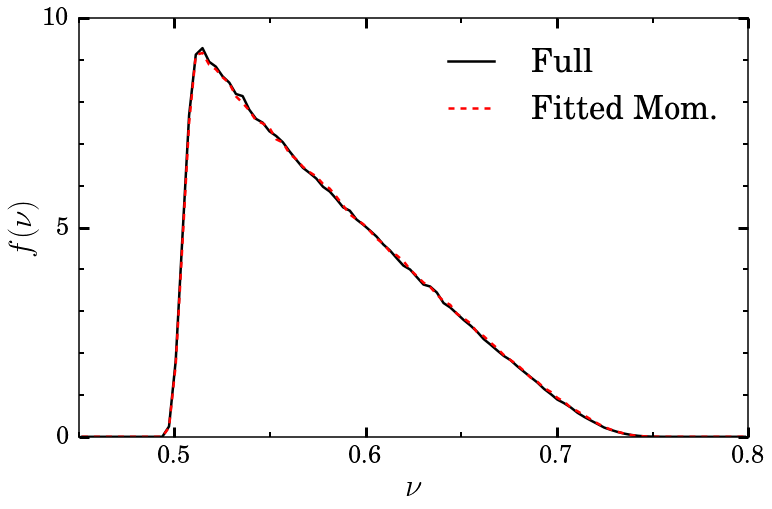}
\caption{Same as Fig.~\ref{fig:scatSpec}, but for a spherical shell $2.5<r<5$ with $\eta =1$, $\kappa_a=1$, and $\kappa_s=100$ around a non-spinning BH. We only show our preferred method for the determination of the momentum of packets, and the result of an evolution without using any approximation in high-scattering regions. Both methods return very similar spectra.} 
\label{fig:scatBHSpec}
\end{figure}

As a more demanding test of our approximate method, we now consider a shell $2.5<r<5$ with $\kappa_s=100$, $\eta=1$, and $\kappa_a=1$, set around a non-spinning black hole of unit mass. 
We perform the simulation in octant symmetry, with a box of length $L_{\rm box} = 7$, a grid spacing $\Delta x \sim 0.13$, and a time step $\Delta t = 0.05$.
The fluid is considered to be at rest with respect to an inertial observer, and thus has a non-zero infall velocity in the grid frame. Accordingly, this test involves non-zero fluid velocity, and a non-trivial metric. It tests, among other things, that our approximate method properly captures the gravitational redshift of the diffusing packets: most of the difference between the energy at which the packets are emitted in the fluid rest frame ($\nu=1$) and the energy measured as they leave the computational domain is due to the gravitational redshift as packets move out of the gravitational potential of the black hole. Figs.~\ref{fig:scatBHLum}-\ref{fig:scatBHSpec} show the flux and energy spectrum of MC packets escaping the domain. We see very good agreement in the spectrum between the full treatment of scatterings and the approximate method, and errors of $\sim 2\%$ in the number of packets leaving the grid. For such a complex setup, we consider this to be an acceptable error at this point.

\subsection{Radiating Spheres}
\label{sec:movingsph}

\begin{figure}
\includegraphics*[width=0.49\textwidth]{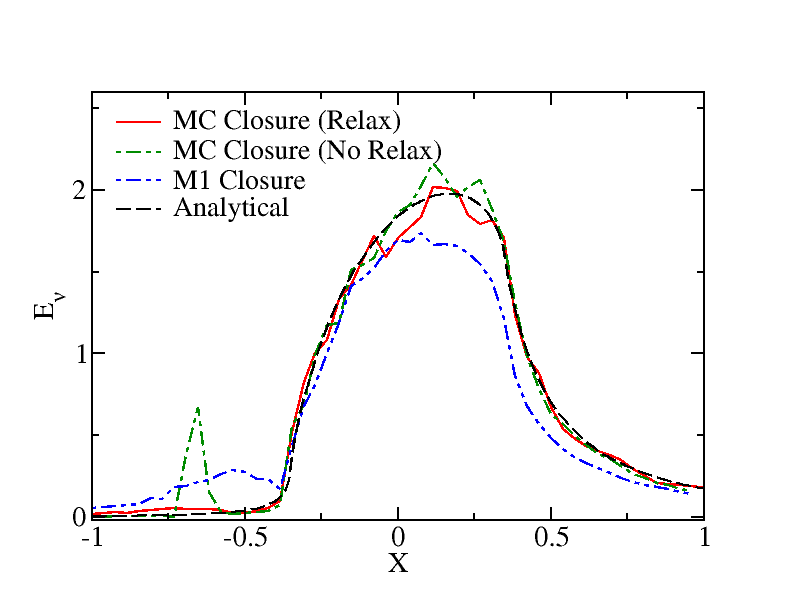}
\caption{Energy density of neutrinos around an emitting sphere of radius $r_s=0.5$ (in its rest frame) moving with a Lorentz factor $W=\sqrt{2}$. We plot $E$ along the axis parallel to the direction of propagation of the sphere and passing through its center (note that due to the boost, the emitting region on this plot is only $-0.35<x<0.35$), at time $t=10GM/c^3$. The dashed black curve shows the known analytical solution for this problem. The dot-dashed blue curve is for the moment formalism with M1 closure. The green dot-dashed curve shows an evolution with the MC-Moment scheme in which we use directly the Eddington tensor computed from the MC evolution. The solid red line uses a weighted average of the MC and M1 closures. In this
problem, the M1 closure does not converge to the correct solution. The MC closure is much better everywhere, except for large errors in a small region behind the emitting sphere. 
The mixed closure provides the best match to the analytical solution.}
\label{fig:MovingSphere}
\end{figure}

As a last idealized test, we consider boosted radiating spheres on a comoving grid. That is, we consider spheres of radius $r_s=0.5$ in which $\eta=\kappa_a=1$ and $\kappa_s=0$, then perform the coordinate transformation $t' = W (t-vx)$, $x'=x/W$, $y'=y$, $z'=z$, with $W=1/\sqrt{1-v^2}$ the Lorentz factor. In that frame, the emitting sphere is boosted, but the center of the sphere is comoving with the computational grid. The 3-metric is $\gamma_{ij} = \delta_{ij}$, the lapse $\alpha=1$, and the shift $\beta^i=(v,0,0)$. This provides us with a good test of semi-transparent systems and Lorentz boosts. We consider $v=0.1$ and $v=1/\sqrt{2}$, a grid spacing $\Delta x = 1/26$, a time step $\Delta t = 0.3\Delta x/W$, and a cubical computational domain of 
size $L=2$ centered on the emitting region. The parameters of the MC evolution are $N_0=100$, $t_d=1$, and $E_p=1e-6$ ($\sim 50$ packets per cell at the center of the emitting
region for $v=1/\sqrt{2}$). When running higher resolution simulations, we use $\Delta x \sim 1/52$, $N_0=200$, and $E_p=6.25e-8$.

The larger velocity case is maybe the most interesting, because it causes problems both for the standard moment scheme with M1 closure and for our MC-Moments algorithm. We show the energy density of neutrinos for that test in Fig.~\ref{fig:MovingSphere}, for simulations using the M1 closure, the pure MC closure, and the mixed MC-M1 closure 
described in Sec.~\ref{sec:MCM1Coupling}. We see that the M1 closure performs surprisingly poorly in this test. The pure MC closure is also far from perfect: it encounters some issues downstream of the emitting sphere, explored in more detail below. This is in fact the only test for which we find that a straightforward use of the MC closure produces visible numerical artifacts (for spheres boosted at a lower velocity, the MC closure performs very well). 

\begin{figure*}
\includegraphics*[width=0.95\textwidth]{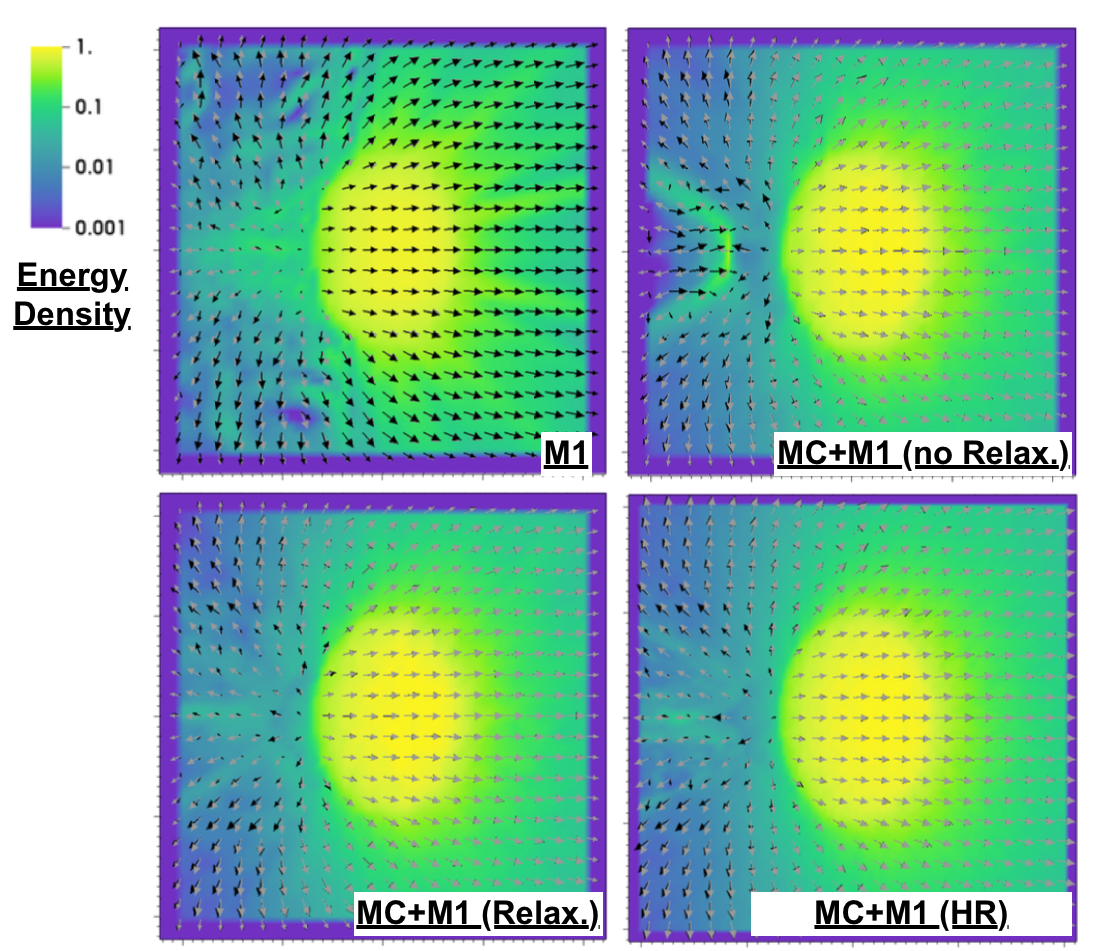}
\caption{Energy density (color) and normalized flux density (arrows, $F_i/E$) of neutrinos around an emitting sphere of radius $r_s=0.5$ moving with a Lorentz factor $W=\sqrt{2}$ along the horizontal axis, at time $t=10GM/c^3$. In each panel, we show the flux in the moment evolution (black), and the flux inferred from the MC packets (grey). The neutrino pressure tensor is computed using a different method in each of the first three panels. {\it Top Left}: from the analytical M1 closure; {\it Top Right}:  from the pressure of the MC packets; and {\it Bottom Left}: from the pressure of the MC packets, with added relaxation towards the predictions of the M1 closure in regions in which the M1 and MC fluxes disagree (i.e. behind the emitting sphere). In the {\it Bottom Right} panel, we show a simulation with half the grid spacing, and 8 times as many MC packets as in the bottom left panel. The M1 simulation shows noticeable artifacts both in front of and behind the emitting sphere. All simulations using the MC closure offer excellent results (in good agreement with the known analytical solution) in front of the sphere. Behind the sphere, large errors are observed if we do not use any relaxation scheme, and the solution remains noisy even with relaxation turned on. Increasing the resolution slightly increases the size of the region in which the solution is accurate (and the M1 and MC fluxes agree), but the solution immediately behind the sphere remains out of the convergent regime. The MC fluxes (grey arrows) are, in all simulations, in good agreement with the exact solution.}
\label{fig:MovingSphereE}
\end{figure*}

\begin{figure*}
\includegraphics*[width=0.95\textwidth]{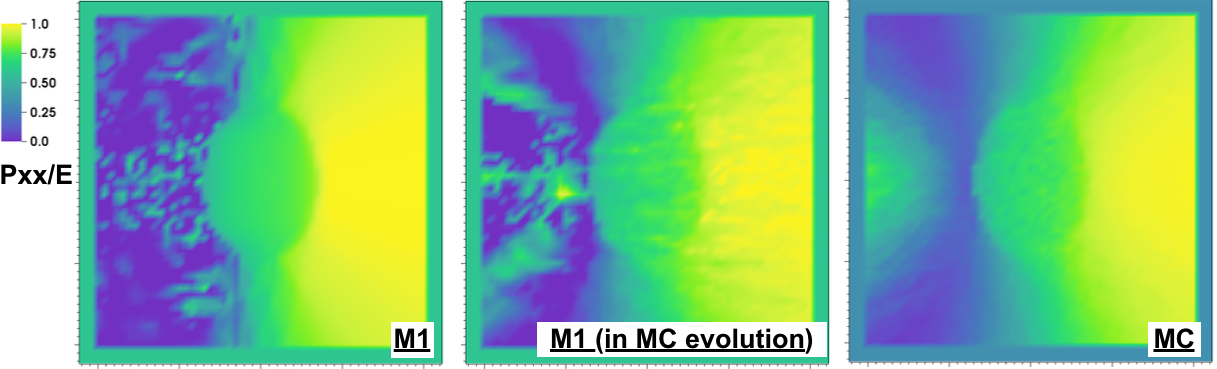}
\caption{Component $P_{xx}/E$ of the neutrino pressure tensor for 3 different choices of closure, for the same simulations as in Fig.~\ref{fig:MovingSphereE}.
{\it Left}: M1 closure; {\it Center}: M1 Closure, computed within a simulation using the MC packets to close the evolution equations; {\it Right}: MC closure, in the same simulation. The M1 closure overestimates $P_{xx}$ in front of the sphere, and generally provides very noisy results behind the sphere. The MC closure shows the expected statistical noise, but is otherwise well-behaved.}
\label{fig:MovingSphereP}
\end{figure*}

To understand the source of these issues, it is useful to look at moments of the neutrino distribution function in the $z=0$ plane. Fig.~\ref{fig:MovingSphereE} shows $E$ and $F_i/E$
(in both the MC and M1 schemes) at time $t=10GM/c^3$. Fig.~\ref{fig:MovingSphereP} shows the Eddington tensor $P_{xx}/E$ at the same time. The pure M1 evolution has 
(at least) two distinct problems: a number of radiation shocks in front of the emitting regions, and a highly inaccurate, noisy evolution behind the emitting region. The simulations
using a MC closure all provide very good results in front of the emitting region. Behind the emitting region, the results are, on average, at most as bad as the M1 solution. Yet, a clear numerical
artifact is observed with the pure MC closure, while convergence towards the correct solution is slow or inexistent when using the mixed closure (see high resolution results on Fig.~\ref{fig:MovingSphereE}). 

The pressure tensor allows us to
understand the source of this artifact. For cell faces orthogonal to the x-axis, the moment equations with the MC closure have 4 characteristic fields with speeds $(-v-\sqrt{\pi_{xx}},-v,-v,-v+\sqrt{\pi_{xx}})$. Thus, for $\pi_{xx} < v^2 =0.5$, all modes propagate to the left. For $\pi_{xx}>0.5$, one mode (coupling $E$ and $F_x$) propagates to the right. 
The erroneous
accumulation of energy observed when using the pure MC closure is located just to the right of the $\pi_{xx}=0.5$ surface, behind the emitting region. This occurs because
the mode with velocity $-v+\sqrt{\pi_{xx}}$ propagates towards that surface, on both sides of it. The $\pi_{xx}=0.5$ surface acts as a critical surface for the neutrino flow, akin to sonic points
in accretion / wind problems. At the point at which that surface crosses the X-axis, we can (nearly) trap radiation energy: $\pi_{yy}$ and $\pi_{zz}$ grow as we move away from the axis, so that a peak in $E$ on the axis can still lead to a flat profile of $P_{yy},P_{zz}$, and thus a quasi-equilibrium solution with $F_y=F_z=0, E\neq0$. It is not, however, a true equilibrium, as energy still flows (slowly) along the $\pi_{xx}=0.5$ surface. 
The M1 closure does not have that problem: a point where the energy density $E$ accumulates is treated as optically thick, and the additional energy  
is then advected with the flow (with some additional diffusion). The mixed closure also avoids that issue, because the M1 and MC fluxes are in strong disagreements behind the 
sphere, and the mixed closure falls back onto the M1 closure in that region (while using the MC closure in front of the emitting sphere).

\begin{figure*}
\includegraphics*[width=0.75\textwidth]{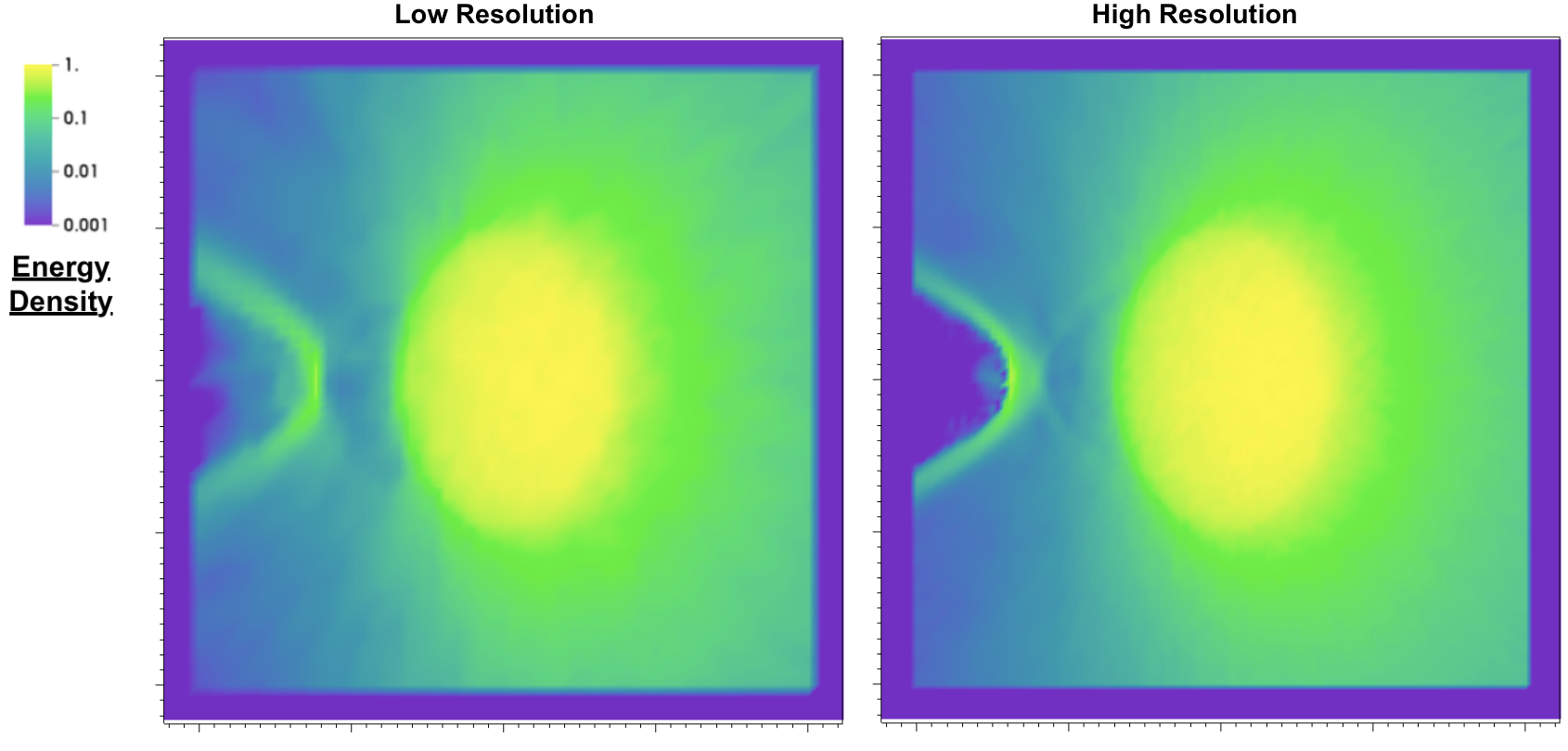}
\caption{Same as Fig.~\ref{fig:MovingSphereE}, but showing two different resolutions, both using the pure MC closure. The shock front nature of the numerical artifact
behind the emitting sphere becomes more visible at higher resolution.}
\label{fig:MovingSphereC}
\end{figure*}

Noticing the existence of that critical surface helps us understand the resolution dependence of the solution for the pure MC closure, 
shown on Fig.~\ref{fig:MovingSphereC}. The high-resolution solution 
improves on the low-resolution solution, except right along the critical surface. This behavior is typical of either shocks or critical points in hydrodynamics simulations: the solution
converges in a global sense (e.g. using the $L_2$ norm), but  not locally (e.g. in the supremum of the error). Or, stated slightly differently, the worse pointwise error does not improve
with resolution, but the size of the region in which that error is localized decreases with the grid spacing. An important open question, that we do not attempt to answer here, is
whether this is sufficient for astrophysical applications -- or, if not, whether the mixed closure used here is an acceptable alternative, or a more advanced coupling scheme has to be
designed. For our target problems, i.e. neutron star-neutron star and black hole-neutron star mergers, a critical surface could be encountered right before merger, when using 
numerical grids comoving with the compact objects. On the other hand, the more interesting post-merger evolution is generally studied with static grids that will not be as sensitive to
this issue. Even in the presence of a critical surface, the MC closure performs better than the existing M1 closure. The main failing of the MC closure is that the error is
concentrated in a small region, and visually obvious. In this idealized test, it is also possible to obtain a more obviously convergent solution by placing the outer boundary in the negative X direction at 
$x\sim-0.5$, thus placing the critical surface out of the computational domain. This is not, however, a luxury that we will have in astrophysical simulations, and thus not an acceptable 
solution. 

\begin{figure}
\includegraphics*[width=0.49\textwidth]{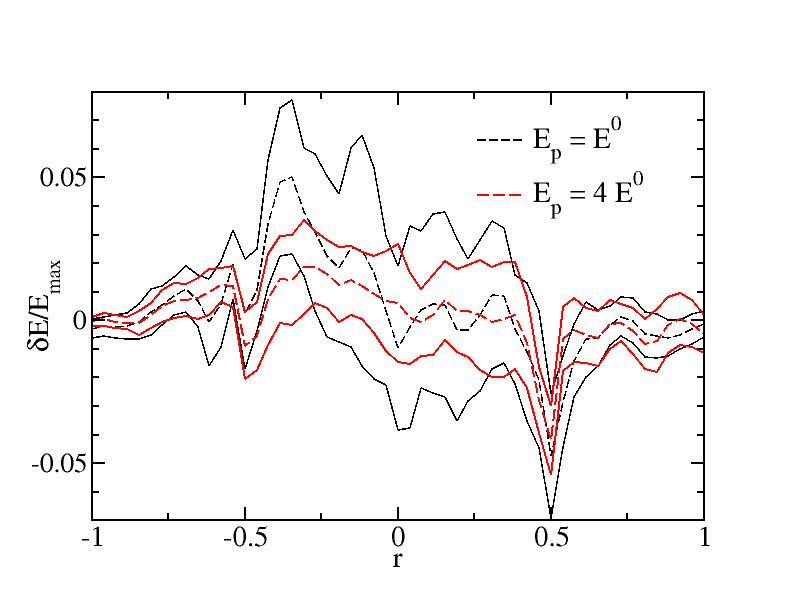}
\caption{Error in the energy density of the neutrinos around an emitting sphere of radius $r_s=0.5$ with a Lorentz factor $W=\sqrt{1.01}$, shown along the axis parallel to the direction of propagation of the sphere and passing through its center. The errors are normalized to the energy density at the center of the sphere. The thinner black curves and thicker red curves show results with a different number of MC packets, but the same resolution: the red curves have four times as many packets. Solid curves show $1-\sigma$ deviations and dashed curves time-averaged values. We average over 50 simulation snapshots in the interval $t=[5,10]$. The errors are relatively small, and using more packets provides the expected slow decrease in the statistical noise due to the use of MC methods.}
\label{fig:ConvEp}
\end{figure}

The lower velocity case, with $v=0.1$, does not show the same dramatic errors as the large velocity case (there is no critical surface for $v=0.1$, as $P_{xx}>0.01$ everywhere). As a less pathological case, it is useful to get a better handle on the accuracy and convergence properties of the code. We first consider two evolutions with the same grid spacing $\Delta x=0.05$, but different number of packets. The first simulation uses $\sim 20$ packets per grid cell at the peak of the energy distribution, while the second uses $\sim 80$ packets per cell. The moments are averaged over $N_0=75$ and $N_0=300$ packets, and the two simulations thus use the same effective averaging timescale for the computation of the MC moments. We use the mixed M1-MC closure, which performs slightly worse than the true MC closure on this problem. The relative error in the average solution is shown on Fig.~\ref{fig:ConvEp}, together with the standard deviation in that solution. Averages are taken over 50 snapshots spaced by $\Delta t=0.1$, starting at $t=5$. In the region with the largest standard deviation, increasing the number of packets by a factor of $4$ decreases statistical variations by roughly the expected factor of $2$. Away from that region, the improvement is significantly slower. While there is a slight improvement in the average solution when increasing the number of packets, particularly at the edge of the emitting sphere, there is no reason to expect that the evolution should converge to the correct solution as the number of packets increases, as we keep the grid spacing constant.

\begin{figure}
\includegraphics*[width=0.49\textwidth]{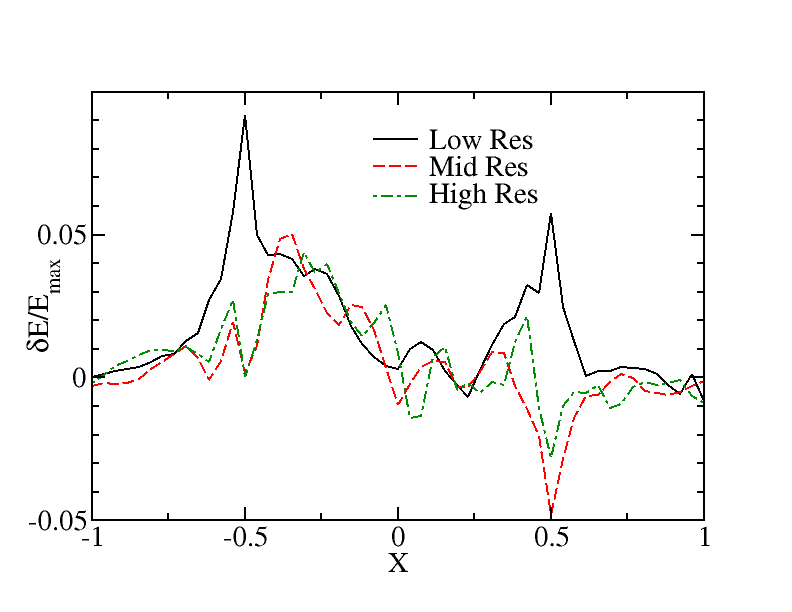}
\caption{Error in the energy density of the neutrinos around an emitting sphere of radius $r_s=0.5$ moving with a Lorentz factor $W=\sqrt{1.01}$, shown along the axis parallel to the direction of propagation of the sphere and passing through its center. The errors are normalized to the energy density at the center of the sphere. We show results for three resolutions, with grid spacing varying by a factor of $1.5$ between each resolution. The number of MC packets on the grid scales as $(\Delta x)^{-3}$. While the solution certainly improves with resolution, noise from the MC evolution makes it impossible to derive an effective order of convergence.}
\label{fig:ConvRes}
\end{figure}

We then consider 3 simulations with different resolutions. The grid spacings are $\Delta x = (0.05*1.5,0.05,0.05/1.5)$, and we vary the number of packets on the grid by scaling $E_p\propto (\Delta x)^{-3}$, which allows us to use $N_0 \propto (\Delta x)^{-1}$ while keeping the effective averaging timescale for the MC moments constant (we use $N_0=75$ for the medium resolution). This does not significantly reduce the statistical noise as the grid spacing decreases, but is fairly typical of the changes in packet number that we may be able to afford in astrophysical simulations. Relative errors for the average energy density, computed as for the previous test, are shown on Fig.~\ref{fig:ConvRes}. We see that the solution improves with resolution, albeit quite slowly. Given the significant noise introduced by the MC evolution, which is at least comparable to the error due to finite grid spacing, a more rigorous study of the convergence of the solution would be a very challenging task.  This points to a potential drawback of our algorithm: true convergence studies would require simulations with unrealistically large differences in the required computational resources. We will instead, at first at least, have to rely on the tests presented here, together with simulations varying a single parameter of the algorithm (e.g. $E_p$, $N_0$,...) to estimate errors.

\subsection{Core-Collapse profile}

\begin{figure}
\includegraphics*[width=0.49\textwidth]{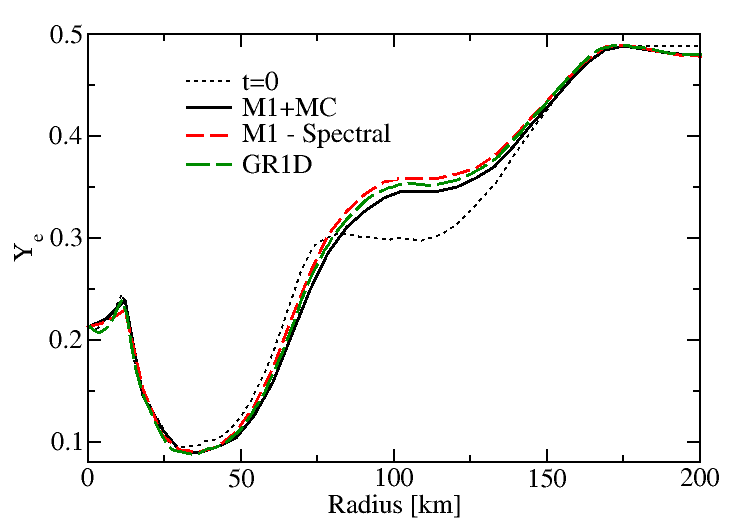}
\caption{Composition of the fluid along the $x^2=x^3=0$ axis, $8\,{\rm ms}$ into the evolution of a post-bounce core-collapse profile. We show results for our mixed MC-Moments algorithm, for our spectral M1 algorithm, and for the GR1D code. We note that {\it grey} M1 evolutions can lead to a wide range of answers for this problem, depending on the method
used to estimate the energy spectrum of the neutrinos~\citep{FoucartM1:2015}.}
\label{fig:YeCollapse}
\end{figure}

Finally, we consider a test of the coupling of the MC-Moments evolution with the evolution of the temperature and composition of the fluid. We use a much more advanced setup, in which we evolve in 3D the post-bounce remnant of a core-collapse simulation from~\cite{Ott2006b}. The result of the simulation is averaged into a 1D, spherically symmetric profile. In our test, we do not evolve the general relativistic equations of hydrodynamics, but modify the temperature and composition of the fluid as neutrinos are emitted and absorbed (the velocity of the fluid remains zero, and the metric is Minkowski). The same test has already been used by~\cite{Abdikamalov2012}, and to test our moment scheme with M1 closure~\citep{FoucartM1:2015,FoucartM1:2016b}. We use the equation of state from~\cite{Shen1998}, with 12 energy groups for the tabulated values of the neutrino emissivities and absorption coefficients. We also use a very coarse resolution $\Delta x \sim 6\,{\rm km}$, and neutrino packets of $E_p \sim 1.8 \times 10^{43}\,{\rm ergs}$. This results in $\sim 500-1000$ packets of each neutrino species leaving the computational grid over a time $\Delta t = GM_\odot/c^3 \sim 5\,\mu{\rm s}$, and $\sim 10^6$ packets per species on the grid at any given time. The minimum averaging timescale used for the computation of the moments is $\tau_d = 2 (1+0.1[r/r_g]) t_g$, with $r_g = GM_\odot/c^2$. 

The region close to the origin has high scattering and absorption opacities, and the heavy-lepton neutrinos have a large region with $\kappa_s \gg 1$, $\kappa_a\ll1$. This test thus provides an interesting way to check whether the interface between optically thick regions and optically thin regions, which are treated differently in our MC-Moments scheme, shows any unphysical artifacts. We can also test that the evolution of the composition and temperature is consistent with previous simulations of this problem, and that the flux of MC packets leaving the grid is consistent with the outgoing flux in the moment evolution in a relatively complex setting.

We have verified that indeed, even in this coupled problem, the evolution of the MC and moment equations remain consistent, in the sense that the neutrino fluxes across the grid boundary are consistent within $\sim 10\%$ between the MC and moment evolutions. We also show on Fig.~\ref{fig:YeCollapse} the composition of the fluid $8\,{\rm ms}$ into the simulation. We compare the results of our new algorithm with an energy-dependent moment scheme using the M1 closure, and with a 1D simulation using the GR1D code~\citep{OConnor2010}, which also uses an energy-dependent moment scheme with M1 closure, and has been compared with full transport simulations. We see good agreements between all the schemes, even though the evolution of the composition is known to be quite sensitive to small differences in the implementation of the transport equations (see e.g. the large variations between different iterations of the gray M1 scheme found in~\citealt{FoucartM1:2015}). The interface between the optically thick and optically thin regions does not cause any problem, as long as the time-averaging of the moments occurs on short timescales in regions with $\kappa_a \gtrsim 1$. This is guaranteed by our current scheme, as our damping timescale scales as the inverse of the number of packets crossing a cell. If we use instead a constant averaging timescale across the entire simulation, and that timescale is too long in optically thick regions (i.e. if $\tau_d \gtrsim \kappa_a^{-1}$), we find that the evolution of the composition in these regions is numerically unstable. On this test problem,
the MC closure does not perform obviously better or worse than the M1 closure, but this is as expected: in spherically symmetric core-collapse simulations, an energy-dependent M1
scheme is already a good approximation to the full transport problem~\citep{Richers:2017}. While many of the other tests devised in this manuscript are designed to study regimes 
in which the M1 closure performs poorly, the main objective of this test is to check that our MC-Moments algorithm can reliably evolve a system with regions of very high and very low
opacities, scattering-dominated atmospheres, and a strong coupling to the evolution of the temperature and composition of the fluid.

We find that for this test problem, the energy-dependent moment formalism with M1 closure agrees with our more advanced transport scheme within $\sim (10-20)\%$ for the luminosity and average energy of the neutrinos, except for the luminosity of heavy-lepton neutrinos which is a factor of $\sim 1.6$ larger when using the M1 closure. As the energy spectrum of heavy-lepton neutrinos is consistent between simulations, the MC and moment evolutions provide consistent results within our simulation, and the evolution of the temperature is also in good agreement in all three simulations, it is unclear at this point what differences in the setup of the test may explain this. The approximate treatment of high-$\kappa_s$ regions would be a natural explanation for discrepancies in the heavy-lepton luminosity, but, if wrong, it should affect the neutrino spectrum and cause disagreements between the neutrino fluxes measured in the moments and MC algorithms. An other indication that these regions are not to blame is that we performed a simulation in which we treated scatterings exactly down to $\kappa_s \Delta \tau = 10$ and evolved packets down to opacities $\kappa_{\rm crit}=20$. That simulation was consistent with the results of our standard configuration, within a few percents. Decreasing the grid spacing by a factor of $2$ led to differences at the $\sim 10\%$ level only. Considering that the high-$\kappa_s$ regions which are so important in setting the characteristics of the heavy-lepton neutrinos are also particularly difficult to capture for all schemes involved, it is likely that only a more dedicated comparison with a full transport code would allow us to determine what difference in the setup of the test explains this one remaining discrepancy.

Overall, this test provides encouraging indications that our mixed MC-Moments scheme can be used in the simulation of astrophysical systems. We also note that, even if the MC evolution is more expensive than the gray moment scheme, at the small number of packets used in this simulation its cost is comparable to the energy-dependent moment scheme! This is despite the fact that the main drawback of the energy-dependent moment scheme, i.e. the potential presence of large energy fluxes in energy space, is not tested by this evolution in which the fluid velocity vanishes everywhere, and the metric is Minkowski. Additionally, nearly all MC packets are owned by the processor on which we evolve the region closest to the center of the neutron star. With a more optimized parallelization of MC packets, the coupled MC-Moments algorithm would presumably be cheaper than the energy-dependent moment scheme in this specific problem and for the low-resolution MC evolution used here.

\section{Conclusions and Prospects}
\label{sec:conclu}

In this work, we show that the evolution of the general relativistic equations of radiation transport using the moment formalism can be complemented by a low-accuracy Monte-Carlo evolution, from which any information required to close the evolution equations of the moments can be extracted. In our algorithm, it is sufficient to evolve MC packets in regions of low absorption optical depth, which avoids many of the more complex and costly steps of MC radiation transport. We also use time-averaged information from the MC evolution to close the evolution equations for the moments, thus reducing at will the number of MC packets necessary to obtain accurate closure relations. We implement this algorithm in the general relativistic hydrodynamics code SpEC, and test our implementation on both simple problems demonstrating that the code properly captures the equations of radiation transport, and on a more complex test coupling the evolution of the radiation to the temperature and composition of the fluid. For this first study of the coupled MC-Moments algorithm, we do not consider full coupling to the evolution of the metric and/or the fluid equations.

Our tests indicate that the coupled MC-Moments evolution always performs better than the more standard approximate M1 closure, and properly reproduces known analytical solutions even for configurations in which the M1 closure fails spectacularly. The coupling of the two systems of equations has to be treated carefully, in order to avoid the growth of numerical instabilities. Accordingly, we provide a detailed description of the steps taken in our current implementation to avoid such instabilities, and of the tests for which simple methods fail. We find that one particular issue that may deserve further investigation is the potential appearance in the simulation of critical surfaces, similar to transitions between supersonic and subsonic flows, where large errors can be encountered (and the solution does not converge if convergence is defined using the supremum norm). While we provide an alternative closure method which sidesteps this issue, the lower performance of that closure on most other tests considered in this manuscript leaves its practical usefulness in more complex systems as an interesting open question. 

We also show that regions of high-scattering opacities can be treated approximately, to avoid the large cost of propagating MC packets through many scattering events, but only as long as the momenta of neutrino packets at the end of this approximate evolution are drawn from a distribution fitted to the result of more detailed scattering experiments. While our method to choose the momentum of the packets introduces additional complexity in the MC algorithm, it does not significantly alter the cost of the evolution.

The coupled MC-Moments algorithm has a number of potential applications in the study of astrophysical systems in which neutrino and photon transport play an important role. It can be used either to obtain the distribution function of neutrinos/photons in time-independent snapshots of a more complex general relativistic hydrodynamics simulation, or can be directly coupled to such evolution. The time averaging of the moments, and the flexibility in the choice of regions within which MC packets are evolved (as opposed to regions in which we assume a thermal distribution of neutrinos/photons when computing closure relations) make it fairly easy to adapt the level of accuracy of the transport scheme to the computational resources available for any given simulation. But its main advantage may be that it actually converges, albeit slowly, to a true solution of Boltzmann's equation. It can also provide a first step towards more expensive, purely MC schemes.

We now expect to use our algorithm first on time-independent snapshots of neutron star merger simulations and then, as computationally possible, in merger simulations in which our MC-Moments scheme is fully coupled to the evolution. There is a lot of information to glean at each of these steps about the impact of neutrinos on the observable properties of mergers. Even evolutions of neutrinos on time-independent simulation snapshots would provide greatly improved estimates of the amount of energy deposition in low-density regions by $\nu \bar\nu$ pair annihilation, as well as a new method to test the accuracy of more approximate schemes and to extract reliable information about the spectrum of neutrinos and their detailed distribution function. The latter may be particularly useful in the study of neutrino oscillations in  neutron star mergers. 

Fully coupled evolutions would, of course, be even more useful. Large uncertainties in the composition of the outflows produced by neutron star mergers, which is set by neutrino-matter interactions, remain a significant roadblock in the production of reliable models for the electromagnetic transients powered by these mergers and for the outcome of r-process nucleosynthesis in that ejecta. Additionally, the deposition of energy by $\nu \bar\nu$ pair annihilation feeds back on the evolution of the matter, and could impact the ability of neutron star mergers to power short gamma-ray bursts. 

While moving from the relatively idealized test problems presented in this work to merger simulations is not a simple task, particularly if one wants to make sure that the computations required by the MC algorithm are optimally distributed on the available computational resources, our results are quite encouraging for coupled MC-Moments algorithms. The radiation transport scheme presented here does not require any modification of the general relativistic hydrodynamics code currently used by SpEC, so that most of the impact of the use of a new radiation scheme has in fact been tested in this work. The work required to move to more realistic astrophysical systems is dominated by the necessity to optimize the implementation of the algorithm, and by the determination of the level of approximation that remains acceptable in the evolution of MC packets in more complex systems: which regions of the simulation can be ignored by the MC algorithm, and how long of an averaging timescale can we use for the moments without creating unphysical artifacts in the simulation. In particular, we note that the parameters affecting the time-averaging of the moments ($N_0, t_d$) have only a minor impact on the quality of the solution in the tests presented here. This may be different in systems with a less trivial time-dependence. 

\section*{Acknowledgements}

The author wish to thank Daniel Kasen and Jennifer Barnes for useful discussions about the use of Monte-Carlo methods in radiation hydrodynamics simulations, as well as Eliot Quataert, Matthew Duez, Sherwood Richers, Ben Ryan, and an anonymous reviewer for their comments and suggestions. The author also wish to thank the attendees of the MICRA conference at Michigan State University and of the INT workshop ``Electromagnetic Signatures of R-process Nucleosynthesis in Neutron Star Binary Mergers'' for stimulating discussions related to this work. Support for this work was provided by NASA through Einstein
Postdoctoral Fellowship grants numbered PF4-150122 awarded by the Chandra X-ray Center, which is operated by the Smithsonian Astrophysical Observatory for NASA under contract NAS8-03060. Computations were performed on the Zwicky cluster at Caltech, supported by the Sherman Fairchild Foundation
and by NSF award PHY-0960291.


\begin{table}
\begin{center}
\caption{Summary of notations used in this manuscript}
{
\begin{tabular}{c|c}
\hline
& {\bf Coordinates}\\
\hline
$(t,x^i)$ & Grid/inertial coordinates \\
$(t',x^{i'})$ & Coordinates in the fluid rest frame \\
$e^\mu_{(\lambda')}$ & Orthonormal tetrad of the fluid rest frame\\
& {\bf Note}: primed indices correspond to quantities in the fluid rest frame\\
\hline
& {\bf Metric}\\
\hline
$g_{\mu\nu}$ & 4-metric \\
$n^\mu$ & unit normal to a constant-$t$ slice\\  
$\alpha$ & Lapse (3+1 decomposition)\\
$\beta^i$ & Shift (3+1 decomposition)\\
$\gamma_{ij}$ & 3-metric (3+1 decomposition)\\
$K_{\mu\nu}$ & Extrinsic curvature (3+1 decomposition)\\
$g,\gamma$ & Determinant of $g_{\mu\nu}$, $\gamma_{ij}$\\
\hline
& {\bf Neutrinos (moments)}\\
\hline
$J$ & Energy density of neutrinos for a comoving observer (0$^{\rm th}$ moment)\\
$H^\mu$ & Flux density of neutrinos for a comoving observer (1$^{\rm st}$ moment)\\
$S^{\mu\nu}$ & Pressure tensor of neutrinos for a comoving observer (2$^{\rm nd}$ moment)\\
$E$ & Energy density of neutrinos for an inertial observer (0$^{\rm th}$ moment)\\
$F^\mu$ & Flux density of neutrinos for an inertial observer (1$^{\rm st}$ moment)\\
$P^{\mu\nu}$ & Pressure tensor of neutrinos for an inertial observer (2$^{\rm nd}$ moment)\\
$\pi_{ij}$ & Eddington tensor, $\pi_{ij} = P_{ij}/E$ \\
$S^\alpha$ & Source term in moment equations (collisional processes)\\
$\eta$ & Energy emissivity (energy integrated)\\
$\eta_N$ & Number emissivity (energy integrated)\\
$\kappa_{a}$ & Absorption opacity (weighted by the spectrum)\\
$\kappa_{s}$ & Scattering opacity (weighted by the spectrum)\\
$\kappa_{N}$ & Absorption opacity (for $Y_e$ evolution,  weighted by spectrum)\\
&  {\bf Note}: tilde quantities are multiplied by $\sqrt{\gamma}$\\
\hline
& {\bf Neutrinos (MC)}\\
\hline
$p^\mu$ & 4-momentum of neutrinos / neutrino packets\\
$N_k$ & Number of neutrinos represented by packet $k$\\
$\nu$ & Average energy of the neutrinos in a packet, in the fluid rest frame\\
$\eta_b$ & Total emissivity per unit volume within the energy bin `b' \\
$\kappa_{a}(\nu)$ & Absorption opacity, at an energy $\nu$\\
$\kappa_{s}(\nu)$ & Scattering opacity, at an energy $\nu$\\
$X_{\rm MC}$ & Quantity $X$ estimated using the MC scheme\\ 
$X_{\rm thick}$ & Contribution to $X$ of optically thick cells with no MC packets evolved\\
$X_{\rm adv}$ & Contribution to $X$ of packets explicitly advected with the fluid\\
$X_{\rm prop}$ & Contribution to $X$ of packets evolved without approximations\\
\hline
& {\bf Time steps in regions of high-scattering}\\
\hline
$\Delta t$ & Time step in inertial coordinates \\
$\Delta t'$ & Time step in the fluid rest frame, $=\Delta t/u^t$ \\
$\Delta t_{\rm fl}$ & Portion of a time step when packets follow the fluid motion\\
$\Delta t_{\rm free}$ & Portion of a time step when packets free-stream\\ 
\hline
& {\bf Fluid}\\
\hline
$\rho$ & Baryon density of the fluid \\
$T$ & Temperature of the fluid\\
$Y_e$ & Electron fraction of the fluid\\
$u^\mu$ & 4-velocity of the fluid\\
$W$ & Lorentz factor of the fluid w.r.t. an inertial observer\\
\hline
& {\bf Free parameters}\\
\hline
$E_p$ & Total energy of a packet in the fluid rest frame (on creation)\\
$t_{\rm diff}$ & Time scale determining when cells use approx. diffusion\\
$\kappa_{\rm crit}$ & Opacity beyond which cells ignore the MC evolution\\
$N_{\rm max}$ & Maximum number of packets created in high-$\kappa$ boundary cells\\
$N_0$ & Desired number of packets used to compute MC moments \\
$t_d$ & Maximum time scale over which we average MC moments\\ 
\end{tabular}
\label{tab:notations}
}
\end{center}
\end{table}



\bibliographystyle{mnras}
\bibliography{PaperRefs} 


\bsp	
\label{lastpage}
\end{document}